 \documentclass[prc,amsmath,amsfonts,showpacs,eqsecnum,floatfix,twocolumn]{revtex4}
 \usepackage{graphicx}  
 \usepackage{bm}  
 \usepackage{pstricks}

\newcommand{\beq}{\begin{eqnarray}}
\newcommand{\eeq}{\end{eqnarray}}

\begin{document}

\title{Nucleon-quark mixed matter and neutron star EOS}

\author{Y.\ Yamamoto$^{1}$}
\email{yamamoto@tsuru.ac.jp}
\author{N.\ Yasutake$^{2}$}
\author{Th.A.\ Rijken$^{3}$$^{1}$}
\affiliation{
$^{1}$RIKEN Nishina Center, 2-1 Hirosawa, Wako, 
Saitama 351-0198, Japan\\
Nishina Center for Accelerator-Based Science,
$^{2}$Department of Physics, Chiba Institute of Technology, 2-1-1 Shibazono
Narashino, Chiba 275-0023, Japan\\
$^{3}$IMAPP, Radboud University, Nijmegen, The Netherlands
}


\begin{abstract}
The nucleon-quark mixed matter is defined in the Brueckner-Hartree-Fock framework,
in which quark densities are determined by equilibrium conditions between 
nucleon and quark chemical potentials, and nucleon-quark interactions play 
critical roles for resulting EoSs (equation of state). 
The two models of EoSs are derived from the nucleon-quark mixed matter (NQMM): 
The NQMM-A EoSs are based on the simple assumption that nucleons and free quarks 
occupy their respective Fermi levels and their Fermi spheres overlap from each other.
In NQMM-B EoSs, the quark Fermi repulsion effect is incorporated on the basis of 
quakyonic matter, meaning that the nucleon Fermi levels are 
pushed up from the quark Fermi sphere by the Pauli exclusion principle.
For the NQMM-A EoSs, the neutron-star mass-radius ($MR$) curves are pushed up
above the region of $M \sim 2.1M_\odot$ and $R_{2.1M_\odot}\sim$ 12.5 km indicated 
by the recent observations, as the $qN$ repulsions increase.
For the NQMM-B EoSs, the similar results are obtained by the combined contributions 
from the $qN$ repulsion and the quark Fermi repulsion. In both models of EoSs, 
the important roles of the $qN$ di-quark exchange repulsions are demonstrated 
to reproduce reasonable values of $M_{max}$ and $R_{2.1M_\odot}$. 
\end{abstract}


\maketitle

\parindent 15 pt

\section{Introduction}

The massive neutron stars (NS) with masses over $2M_{\odot}$ have been reliably 
established by the observations of J1614$-$2230 \cite{Demorest10}, 
J0348+0432 \cite{Antoniadis13}, J0740+6620 \cite{Cromartie2020} and
J0952-0607 \cite{Romani2022}. Despite these observations of massive
neutron stars over $2M_{\odot}$, the hyperon mixing in neutron-star matter 
brings about a remarkable softening of the EoS (equation of state) 
and a maximum mass is reduced to a value far less than $2M_{\odot}$, 
being called ``hyperon puzzle in neutron stars".
The EoS softening is caused by changing of high-momentum neutrons
at Fermi surfaces to low-momentum hyperons via strangeness 
non-conserving weak interactions overcoming rest masses of hyperons.
Generally, such mechanisms can be due to possible appearance of
other hadronic degrees of freedom, such as
$\Delta$ isobars \cite{Drago}, meson condensates 
\cite{Kaplan,Brown,Thorsson,Lee,GleSch} or
quark phases \cite{Baldo2006,Ozel,Weissenborn, Klahn, Bonanno,
Lastowiecki2012, Shahrbaf1,Shahrbaf2,Otto2020,KBH2022,YYR2022,YYR2023}.
Since such mechanisms of EoS softening are inevitable in neutron-star matter,
it has been one of the central issues in this field to model EoSs giving
star masses over $2M_{\odot}$.

From the analyses for the X-ray data taken by the 
{\it Neutron Star Interior Composition Explorer} (NICER)
and the X-ray Multi-Mirror (XMM-Newton) observatory,
the radius information of NSs have been obtained for the massive NS PSR
J0740+6620 \cite{Miller2021,Riley2021,Legred2021,Salmi2024,Dittmann2024}.
The radius information of massive NSs give important constraints 
for neutron-star EoSs, which are demonstrated critically
by reproducing the neutron-star radii \cite{YYR2023}.

Recently, they have given a more precise measurement of the radius of
PSR J0740+6620 using updated NICER data as $R= 12.49^{+1.28}_{-0.88}$ km
with the determined mass $M = 2.08 \pm 0.07 M_\odot$ \cite{Salmi2024}.
We adopt their median values of $M=2.1 M_\odot$ and the radius
$R_{2.1M_\odot}= 12.5$ km at $M=2.1 M_\odot$ for comparison with 
our calculated results. These values are used as the criterion for EoSs,
which is far more severe than the one using only the mass values such as 
$2M_{\odot}$. In mass and radius ($MR$) relations of neutron stars, 
the criterion for $MR$ curves is to reach the point 
($M=2.1 M_\odot, R_{2.1M_\odot}=12.5$ km).

There have been proposed several mechanisms to reproduce masses over 
$2M_{\odot}$, solving the ``hyperon puzzle".
Among them, the baryonic approach is to introduce the repulsive 
hyperonic three-body forces at the baryon level 
\cite{NYT,Vidana11,YFYR14,YFYR16,YTTFYR17,Lonardoni,Logoteta,Cerstung}.
In \cite{YFYR14,YFYR16,YTTFYR17}, the multi-pomeron exchange potential 
(MPP) was introduced as a model of universal repulsions among three and 
four baryons on the basis of the extended soft core (ESC) baryon-baryon 
interaction model developed by two of the authors 
(T.R. and Y.Y.) and M.M. Nagels \cite{ESC16I,ESC16II,ESC16III}.
In their modeling for hyperonic three-body repulsions,
the EoS softening by hyperon mixing is not completely recovered by 
the MPP repulsions: The maximum masses do not significantly exceed
$2M_\odot$, even if MPP repulsions are taken strong enough. 
It seems difficult to reproduce both of $M_{max}\sim 2.1M_\odot$ and 
$R_{2.1M_\odot}\sim$ 12.5 km by hadronic-matter EoSs with hyperon mixing,
as discussed in the following section.

The challenging subject is to study quark deconfinement phase transitions 
from a hadronic-matter EoS to a sufficiently stiff quark-matter EoS 
giving $M_{max}\sim 2.1M_\odot$ and $R_{2.1M_\odot}\sim$ 12.5 km.
It is known that quark-hadron phase transitions should be crossover 
or at most of weak first-order in order to obtain EoSs stiff enough,
because strong first-order transitions soften EoSs remarkably 
\cite{Baldo2006,Ozel,Weissenborn,Klahn,Bonanno,Lastowiecki2012,
Shahrbaf1,Shahrbaf2,Otto2020,KBH2022,YYR2022}.
Then, it is essential that repulsive effects in quark phases are needed to 
result in massive stars over $2M_{\odot}$. Without such repulsive effects, 
the quark mixing in neutron-star matter brings about a remarkable softening 
of the EoS and a maximum mass is reduced to a value far less than $2M_{\odot}$, 
where the softening is caused by changing of high-momentum neutrons 
to low-momentum quarks.
In the Nambu-Jona-Lasinio (NJL) model, for instance, repulsions to 
stiffen EoSs are given by vector interactions.
In the case of our quark-hadron transition (QHT) model 
\cite{YYR2022, YYR2023}, the quark-quark ($qq$) repulsions 
composed of meson-exchange and one-gluon-exchange potentials.

Another type of quark phase in neutron-star interiors is given by
the so-called quarkyonic matter
\cite{MP2007,HMP2008,MR2019,HMLCP2019,DHJ2020,ZL2020,MHPC2021,Cao2022,YYR2023},
where the degrees of freedom inside the Fermi sea are treated as free quarks,
and nucleons exist at the surface of the Fermi sea.
In the quarkyonic matter, the existence of free quarks inside
the Fermi sphere gives nucleons extra kinetic energy by pushing them up
to higher momenta, leading to increasing pressure.
This mechanisms stiffening EoSs are completely different from the 
quark-hadron transition models in which the essential roles for EoS 
stiffening are played by the $qq$ repulsions. 
In our previous work \cite{YYR2023}, we investigated the roles of 
quark-quark ($qq$) and quark-neutron ($qn$) interactions in the 
quakyonic matter, which were not taken into account in \cite{MR2019}.
However, there remain two important problems in our treatment: The first 
is that the roughly approximated version is used for the $qn$ interactions. 
The second is that in quakyonic matter formalism \cite{MR2019}
the thickness parameter $\Delta$ for the neutron Fermi layer
plays the decisive role for neutron-star $MR$ (mass-radius) curves, 
and then no important conclusions can be drawn regarding minor 
effects of $qn$/$qq$ interactions.

In this paper, we propose the nucleon-quark mixed matter (NQMM) model 
in the Brueckner-Hartree-Fock (BHF) framework, which is suitable to 
clarify effects of quark-nucleon ($qN$) and $qq$ interactions.
As a first step, in treating mixed-matter hyperon and
$s$-quark mixings are not included for simplicity.

In this framework quark densities in nucleon-quark mixed matter are 
determined by equilibrium conditions between chemical potentials 
of neutrons and free $u$ and $d$ quarks without using ad hoc parameters 
such as the Fermi-layer thickness $\Delta$.
We define here the two models of nucleon-quark mixed matter;
NQMM-A and NQMM-B.
In NQMM-A, nucleons (free quarks) occupies simply their Fermi sphere
from zero momentum to Fermi momentum $k_F^N$ ($k_F^q$).
In NQMM-B, the Fermi repulsion effects for nucleons are incorporated 
on the basis of quakyonic matter, in which nucleon Fermi levels are 
pushed up to those with higher momenta by the Fermi exclusion 
for nucleons from the quark Fermi sphere. 
In the derivations of EoSs for NQMM-A and NQMM-B,
calculations are performed with use of the realistic $qN$ interactions 
given in \cite{Rijken24a,Rijken24b} together with the realistic
$NN$ and $qq$ interactions.
The derived EoSs are used to obtain the neutron-star $MR$ curves 
by solving the Tolmann-Oppenheimer-Volkoff (TOV) equations.
Then, it is possible to study how the conditions of reproducing
$M_{max}\sim 2.1M_\odot$ and $R_{2.1M_\odot}\sim$ 12.5 km
are realized by our EoSs.

This paper is organized as follows:
In Sect.II-A, the hadronic-matter EoSs in our previous works
\cite{YFYR14,YFYR16,YTTFYR17} are recapitulated, which is the basis
of deriving the nucleon-quark mixed matter EoS. 
In Sect.II-B, our $qN$ interactions are explained, which play
important roles for the derived EoSs and neutron-star $MR$ curves.
In Sect.II-C, our EoSs for NQMM-A and NQMM-B are 
formulated in the BHF framework.
In Sect.III-A, the calculated results are shown for densities, 
energy densities, chemical potentials and pressures 
in nucleon-quark mixed matter, leading to our EoSs.
In III-B, the $MR$ curves of neutron stars are obtained 
by solving the TOV equation with our EoSs for NQMM-A and NQMM-B.
The conclusion of this paper is given in Sect.IV.

\section{Nucleon-Quark mixed matter}

\subsection{Toward quark mixing in hadronic matter}

The hadronic matter is defined as $\beta$-stable hyperonic nuclear matter.
As a starting point for investigating the nucleon-quark mixed matter,
we recapitulate a typical hadronic-matter EoS composed of $n$, $p^+$, $\Lambda$, $e^-$
in the BHF framework with use of the ESC baryon-baryon ($B\!B$) interaction model 
\cite{YFYR14,YFYR16,YTTFYR17}.

Our baryonic interactions are composed of two-body part $V_{BB}$ and 
three-body part $V_{BBB}$.
$BB$ G-matrix interactions ${\cal G}_{BB}$ are derived from
$BB$ bare interactions $V_{BB}$ or $V_{BB}+V_{BBB}$ \cite{YFYR14}. 
They are given for each $(BB',T,S,P)$ state, $T$, $S$ and $P$
being isospin, spin and parity in a two-body state, respectively,
and represented as ${\cal G}_{BB'}^{TSP}$.
In the following sections, we need only the 
nucleon-nucleon sectors, ${\cal G}_{NN}^{SP}$.

As is well known, the nuclear-matter EoS is stiff enough to assure neutron-star masses
over $2M_{\odot}$, if the strong three-nucleon repulsion is taken into account.
However, there appears a remarkable softening of EoS by inclusion of exotic degrees 
of freedom such as hyperon mixing. 
As one of the ideas to avoid this ``hyperon puzzle", it was proposed that 
the three-body repulsions worked universally for every kind of baryons \cite{NYT}.
In \cite{YFYR14,YFYR16,YTTFYR17}, 
the multi-pomeron exchange potential (MPP) was introduced as a model of universal 
repulsions among three and four baryons. 
The recent ESC works are mentioned in \cite{ESC16I,ESC16II,ESC16III}. 

In \cite{YTTFYR17} they proposed three versions of MPP (MPa, MPa$^+$, MPb),
where MPa and MPa$^+$ (MPb) include the three- and four-body (only three-body) repulsions.
The obtained $MR$ curves are given in Fig.3 \cite{YTTFYR17}, where
the curves move upwards with increase of MPP repulsions.
The important criterion in this paper is the value of $R_{1.4M_\odot}$: 
We adopt MPb, giving $R_{1.4M_\odot} \approx 12.4$ km.
As shown later, this $R_{1.4M_\odot}$ value of MPb persists in 
the nucleon-quark mixed matter EoS.

In \cite{YTTFYR17}, values of $M_{max}$ and $R_{2M\odot}$ are obtained 
from the EoSs including MPb for $\beta$-stable nuclear matter with 
and without $\Lambda$ mixing, which are $M_{max}/M_\odot$= 2.06 (2.19), and 
$R_{2M_\odot}$= 11.3 km (11.8 km) for the EoS with (without) $\Lambda$ mixing.
The values of $R_{2M_\odot}$ for the hadronic matter EoSs 
are substantially smaller than 12.5 km.
By such hadronic-matter EoSs, it difficult to satisfy the criterion 
reproducing ($M_{max}=2.1M_\odot$, $R_{2.1M_\odot}=12.5$ km).
It is commented that the three-nucleon repulsion included in MPb is stronger 
than the corresponding one (UIX) in the standard model by APR \cite{APR98} 
giving rise to $R_{1.4M_\odot} \approx 11.6$ km \cite{Togashi1}.

\bigskip

In order to explore possibilities of getting larger values of $R_{2.1M_\odot}$
under $M_{max}\sim 2.1M_\odot$, the BHF framework for hadronic matter is 
extended to the nucleon-quark mixed matter composed of $n$, $p$, $u$, $d$ 
and $e^-$, where strangeness degrees of freedom are not taken into account 
for simplicity. Because $u$ and $d$ quarks are treated in the BHF base, 
it is necessary to use quark-nucleon ($qN$) and quark-quark ($qq$) 
two-body interactions in BHF calculations as well as $NN$ interactions. 
In the treatment of quark mixing, we learn from the simple case of $\Lambda$ 
mixing in neutron matter where $\Lambda$ mixing rates are determined by 
chemical equilibrium conditions $\mu_n=\mu_\Lambda$ with chemical potentials 
$\mu_n$ and $\mu_\Lambda$ of neutron and $\Lambda$, respectively.
In the case of neutron-quark mixing, correspondingly, 
quark mixing rates in neutron matter are determined by chemical 
equilibrium conditions $\mu_n=\mu_u+2\mu_d$ with chemical potentials
$\mu_u$ and $\mu_d$ of $u$ and $d$ quarks, respectively.

\subsection{$qN$ and $qq$ interactions}

In BHF calculations of nucleon-quark mixed matter,
quark-quark ($qq$) and quark-nucleon ($qN$) two-body 
interactions are needed as well as $NN$ interactions. 
In our previous works \cite{YYR2022,YYR2023},
the quark-matter calculations were performed with use of 
the two-body $qq$ interaction.
The basic part of this $qq$ interaction is given by the extended
meson-exchange (EME) potential $V_{EME}^{(qq)}$, derived from
the ESC $BB$ potential so that the $qqM$ couplings are related 
to the $BBM$ couplings through folding procedures with Gaussian 
baryonic quark wave functions. In \cite{YYR2022}, this $qq$ potential 
was named as Q0, and the more repulsive versions are given by adding 
instanton-exchange and one-gluon exchange (OGE) potentials (Q1), 
and furthermore the multi-pomeron exchange potential (Q2). 
In this work, we use the simplified version Q3 composed of EME and OGE 
potentials, being adjusted so as to be similar to Q2. Qualitatively, 
however, resulting EoSs do not much depend on whether Q0, Q1, Q2 or Q3 
is used for $qq$ interactions, because quark densities in nucleon-quark 
mixed matter are not large compared to nucleon densities and partial 
pressures of free quarks are far smaller than those of nucleons.

In \cite{Rijken24b}, $qq$ potentials $V_{EME}^{(qq)}$ and $qN$ potentials
$V_{EME}^{(qN)}$ are derived together, in which meson-quark-quark and
meson-nucleon-nucleon couplings are determined consistently.
In addition to the meson-exchange potentials, we introduce the di-quark 
exchange (DQE) potential $V_{DQE}^{(qN)}$ derived in \cite{Rijken24a}: 
The total $qN$ interaction is 
\begin{eqnarray}
V^{(qN)} &=& V_{EME}^{(qN)} + V_{DQE}^{(qN)} \ . 
\end{eqnarray}
As shown later, $V_{DQE}^{(qN)}$ plays a more important role
than $V_{EME}^{(qN)}$ in our calculations.

In Appendix~\ref{app:DQEXCH} a brief description of the derivation of the 
di-quark exchange contact potential is given. This is based on a phenomenological 
description of the confinement-deconfinement at high nucleon-densities via 
a nucleon-triquark $\lambda_3$-coupling, using a quark-diquark description 
of the triquark. The resulting NJL-type quark-nucleon interaction is
%
\begin{eqnarray}
	{\cal L}^{(2)}_{int} &=& -\lambda_3^2 \left(\bar{\psi}(x)\gamma_5\gamma_\mu \bm{\tau} Q\right)
\cdot\left(\bar{Q}\gamma_5\gamma^\mu \bm{\tau}\psi\right)/{\cal M}^2 
\label{eq:Ldeconf3} \end{eqnarray}
with ${\cal M}= \hbar c$. (Choosing ${\cal M}$ differently merely means a rescaling of 
$\lambda_3$.)  In (\ref{eq:Ldeconf3}) we 
used the isospin spinor $Q=(u,d)$.

The di-quark exchange (central) potential from the Lagrangian
(\ref{eq:Ldeconf3})
is given by \cite{Rijken24a}
\begin{eqnarray}
&& V_{DQE}^{(qN)}(r) = -\lambda_3^2 \frac{\Lambda}{4\pi \sqrt \pi}
\frac{\Lambda^2}{{\cal M}^2}\
(\bm{\tau}_1\cdot\bm{\tau}_2)(\bm{\sigma}_1\cdot\bm{\sigma}_2)\ {\cal P}_x\cdot
\nonumber\\
&& \times \left [1-\frac{3\Lambda^2}{4 m_N m_q}
\left(1-\frac{\Lambda^2 r^2}{6} \right) \right]
\exp \left[-\frac{\Lambda^2 r^2}{4} \right] ,
\end{eqnarray}
where the baryon-triquark coupling $\lambda_3$ in MeV$^{-2}$, ${\cal P}_x$ the space-exchange
operator, and $\Lambda$ and ${\cal M}$, can be treated as adjustable parameters.

\bigskip
Considering the chiral symmetry breaking as the QCD non-perturbative effect,
constituent quark masses in quark matter become smaller than those in vacuum 
and move to current masses in the high-density limit. 
In our previous works \cite{YYR2022,YYR2023},
we introduced phenomenologically the density-dependent quark mass
\begin{eqnarray}
M_Q^*(\rho_Q) = M_0/[1+\exp \{\gamma (\rho_Q-\rho_c\}] +m_0 +C
\label{mstar}
\end{eqnarray}
with $C=M_0-M_0/[1+\exp (-\gamma \rho_c)]$ assuring $M_Q^*(0) = M_0+m_0$,
where $\rho_Q$ is quark density.
The effect of using this quark masses is demonstrated in Fig.\ref{MRA},
where the parameters are chosen as $\rho_c=7\rho_0$, $M_0=362$ MeV
and $\gamma=1.2$ \cite{YYR2023}.

\subsection{Framework of Nucleon-Quark mixed matter}
Our nucleon-quark mixed matter is composed of nucleons ($n$ and $p^+$),
quarks ($u$ and $d$) and electrons ($e^-$), where a nucleon number density 
$\rho_N$ is given by a sum of neutron and proton densities, 
$\rho_N=\rho_n + \rho_p$, and a quark number density $\rho_q$ by a sum of 
u-quark and d-quark densities, $\rho_q=\rho_u+\rho_d$.
In our treatment of nucleon-quark mixed matter, the BHF framework is 
adopted on the basis of two-body $NN$, $qq$ and $qN$ potentials. 
Correlations induced by bare potentials are renormalized into 
coordinate-space G-matrix interactions, treated as effective 
two-body interactions used in matter calculations. 
G-matrix interactions ${\cal G}_{NN'}$, ${\cal G}_{qN}$, 
${\cal G}_{qN}$ and ${\cal G}_{qq'}$ with $N,N'=n,p$ and $q,q'=u,d$ 
are derived from the above bare $NN$, $qN$ and $qq$ interactions.
They are given for each $(T,S,P)$ state, $T$, $S$ and $P$
being isospin, spin and parity in a two-body state, respectively.

Single particle potentials of $N$ and $q$ are given by
\begin{eqnarray}
U_N(k)&=&\sum_{N'=n,p} U_{N}^{(N')}(k) +\sum_{q'=u,d} U_{N}^{(q')}(k)
\nonumber\\
&=& \sum_{N'=n,p} \sum_{k'<k_F^{(N')}} 
\langle kk'|{\cal G}_{NN'}|kk'\rangle
\nonumber\\
&& +\sum_{q'=u,d} \sum_{k'<k_F^{(q')}}
\langle kk'|{\cal G}_{Nq'}|kk'\rangle
\label{UNQ} 
\\
U_q(k)&=&\sum_{q'=u,d} U_{q}^{(q')}(k) +\sum_{N'=n,p} U_{q}^{(N')}(k)
\nonumber\\
&=& \sum_{q'=u,d} \sum_{k'<k_F^{q'}} \langle kk'|{\cal G}_{qq'}|kk'\rangle
\nonumber\\
&& +\sum_{N'=n,p} \sum_{k'<k_F^{(N')}}
\langle kk'|{\cal G}_{qN'}|kk'\rangle
\label{UQN} 
\end{eqnarray}
where $k_F^{N}$ and $k_F^{q}$ is the Fermi momenta of 
nucleon $N$ and quark $q$, respectively.
Spin and isospin quantum numbers are implicit.

The quark energy density for $q=u,d$ in our nucleon-quark mixed matter
is given by
\begin{eqnarray}
\varepsilon_q&=& m_q \rho_q +
g_s N_c \int_0^{k_F^{q}} \frac{d^3k}{(2\pi)^3}
\nonumber \\
&& \left\{\sqrt{\hbar^2 k^2+m_{q}^2}+\frac 12 U_{q}(k) \right\} \ .
\label{edenq} 
\end{eqnarray}
where Fermion spin and quark-color degeneracies give rise 
to $g_s=2$ and $N_c=3$, respectively.

The nucleon energy density for $N=n,p$ is given by
\begin{eqnarray}
\varepsilon_N&=& m_N \rho_N + \tau_N + \upsilon_N 
\nonumber \\
&=& m_N \rho_N + g_s \int_{k_0^N}^{k_1^N} \frac{d^3k}{(2\pi)^3}
\nonumber \\
&& \left\{\sqrt{\hbar^2 k^2+m_N^2}+\frac 12 U_N(k)\right\} \ .
\label{edenn} 
\end{eqnarray}
It is necessary, here, to explain upper and lower limits of integral
($k_0^N$ and $k_1^N$). In the simple case of deriving NQMM-A EoSs,
they are chosen as $k_1^N=k_F^N$ and $k_0^N=0$ as usual.
 
On the other hand, in the case of deriving NQMM-B EoSs,
according to the concept of the quarkyonic matter \cite{MR2019}, 
where interacting quarks near the Fermi sea form interacting neutrons, 
and the remaining quarks fill the lowest momenta up to $k_F^q$.
The nucleon Fermi levels of $0<k<k_F^N$ are pushed up to those 
of $k_0^N<k<k_1^N$ as a result that nucleons below $k_0^N$ 
are excluded by the Pauli principle.

Then, $k_0^N$ and $k_1^N$ are given as follows:  
The density of the nucleon Fermi sphere with
radius $k_0^N$ is given by $\frac{(k_0^N)^3}{3\pi^2}$, and
the density of the quark Fermi sphere with radius $k_F^q$ is 
given by $\rho_q=\frac{3 N_c (k_F^q)^3}{3\pi^2}$.
Considering that the $k_0^N$ value is determined by the
condition that the densities of nucleon and quark Fermi spheres
are equal to each other,  we have
\begin{eqnarray}
\frac{(k_0^N)^3}{3\pi^2}=\rho_q \ ,
\label{K0KF} 
\end{eqnarray}
from which the value of $k_0^N$ is obtained. 
The value of $k_1^N$ is obtained from the relation
\begin{eqnarray}
\frac{(k_1^N)^3}{3\pi^2} -\frac{(k_0^N)^3}{3\pi^2}
=\frac{(k_F^N)^3}{3\pi^2}=\rho_N \ .
\label{K1KF} 
\end{eqnarray}
These values of $k_0^N$ and $k_1^N$ are lower and upper limits 
of the integral in Eq.(\ref{edenn}).
The relations (\ref{K0KF}) and (\ref{K1KF}) are considered as 
plausible assumptions to determine $k_0^N$ and $k_1^N$, respectively,
which plays a decisive role to give nucleon partial pressures pushed 
up by the quark Fermi repulsion in the NQMM-B case. In this work, 
it is out of our scope to derive the quark Fermi repulsion 
more microscopically.

Here, the potential-energy part in Eq.(\ref{edenn})
is approximated by the following expression
\begin{eqnarray}
&& {\bar \upsilon_N}= 
 g_s \int_{0}^{k_F^N} \frac{d^3k}{(2\pi)^3}
\left\{\frac 12 U_N(k)\right\} \ ,
\label{vden} 
\end{eqnarray}
meaning that the Fermi repulsion for low-momentum
neutrons are determined by kinetic-energy densities.
Furthermore, this Fermi exclusion effects for low-momentum 
components are neglected for protons.

\bigskip

Our total energy density is given by
\begin{eqnarray}
\varepsilon=\varepsilon_n+\varepsilon_p+\varepsilon_u+\varepsilon_d+\varepsilon_e \ .
\end{eqnarray}
The chemical potential $\mu_i$ ($i=n,p,u,d,e$) and pressure $P$
are expressed as
\begin{eqnarray}
&&\mu_i = \frac{\partial \varepsilon_i}{\partial \rho_i} \ , 
\label{chem} \\
&& P = \sum_{i=n,p,u,d,e} \mu_i \rho_i -\varepsilon \ .
\label{press}
\end{eqnarray}

When $P$ and $\varepsilon$ are given, sound velocities are defined by
$c_s^2= \partial P/\partial \varepsilon$. 
Our BHF framework is basically non-relativistic, and a causal condition 
of $c_s <c$ is not always assured: In regions of $c_s >c$, 
sound velocities are approximated to be $c_s =c$.

\bigskip

In the EoS of $\beta$-stable nucleon-quark mixed matter 
composed of $n$, $p$, $u$, $d$ and $e$,
the equilibrium conditions are given as

\medskip
\noindent
(1) chemical equilibrium conditions,
\begin{eqnarray}
&& \mu_n= \mu_p+\mu_e
\label{eq:c0}
\\
&& \mu_n = \mu_u+2 \mu_d 
\label{eq:c1}
\end{eqnarray}
\noindent
(2) charge neutrality,
\begin{eqnarray}  
    \rho_p = \rho_e 
\label{eq:c2}
\end{eqnarray}
\noindent
(3) baryon number conservation,
\begin{eqnarray}
\rho = \rho_n +\rho_p + \rho_Q \ ,
\label{eq:c3}
\end{eqnarray}
where $\rho_Q$ is a baryonic (3 quarks)  number density
which is related to a quark number density $\rho_q$
by $\rho_Q = \rho_q/3$ with $\rho_q=\rho_u+\rho_d$,
and $\rho$ is a total number density.
For simplicity, the ratio of $u$ and $d$ quarks in nucleon-quark 
mixed matter is assumed to be $\rho_u/\rho_d=1/2$ as well as 
the one in neutron-quark mixed matter, and the equilibrium 
condition $\mu_p =2 \mu_u+ \mu_d$ is not taken into account.

Defining
$Y_p=\rho_p/\rho_N$ with $\rho_N=\rho_n + \rho_p$ and
$Y_Q=\rho_Q/(\rho_N+\rho_Q)$,
energy densities ($\varepsilon_n$, $\varepsilon_p$, $\varepsilon_u$,
$\varepsilon_d$) and chemical potentials ($\mu_n$, $\mu_p$, $\mu_u$, $\mu_d$)
are given as a function of $\rho_N$, $Y_p$ and $Y_Q$.
Then, $Y_p$ and $Y_Q$ values included in the EoS are obtained by solving 
the equations $\mu_n= \mu_p+\mu_e$ (\ref{eq:c0}) and $\mu_n = \mu_u+2 \mu_d$ 
(\ref{eq:c1}).
Solutions are obtained in an approximate way: 
First, $Y_p$ values are obtained by solving the equation $\mu_n= \mu_p+\mu_e$
for given values of $\rho_N$ within nucleon components (taking $Y_Q=0$) 
in nucleon-quark mixed matter. Next, $Y_Q$ values are obtained by
solving the equation $\mu_n = \mu_u+2 \mu_d$ for given values of
$\rho_N$ and $Y_p$.
The radius of the quark Fermi sphere (Fermi momentum) $k_F^q$ 
is related to $Y_Q$ by
\begin{eqnarray}  
    k_F^q = (3\pi^2 Y_Q (\rho_N+\rho_Q))^{1/3} \ .
\label{eq:c4}
\end{eqnarray}
Then, lower and upper limits of the integral in Eq.(\ref{edenn}),
$k_0^N$ and $k_1^N$, are obtained by using (\ref{K0KF}) and (\ref{K1KF}),
meaning that the $Y_Q$ values determine the nucleon Fermi layer
pushed up by the Fermi repulsion.

The quark-mixing rates $Y_Q$ are determined by the equilibrium condition 
Eq.(\ref{eq:c1}) for nucleon and quark chemical potentials.
The latter includes constituent quark masses $m_q$, values of which are
related sensitively to onset densities of quark mixing
in nucleonic matter. A simple choice of constituent quark mass is
$m_q=m_N/3=313$ MeV. In \cite{YasMar}, however, they obtain $m_q=362$ MeV 
which is determined to reproduce $m_N=938$ MeV with the confining
and one-gluon exchange potentials among three quarks.
We adopt their values of 362 MeV for constituent quark masses.
It is confirmed that this large value of 362 MeV is favorable to give 
larger values of quark onset densities than the small value of 313 MeV, 
being important for deriving reasonable $MR$ curves of neutron stars. 
For further tuning of quark onset densities,
we introduce corrections for mass terms of quark chemical potentials
by replacing $m_q$ to $m_q+\Delta m_q$ with parameters $\Delta m_q$.
This tuning of quark onset densities by $m_q$ is important
for obtaining reasonable $MR$ curves.

\bigskip

In the above, our quark states in nucleon-quark mixed matter are
represented in the BHF framework, where formations of e.g. superconducting
pairs are neglected, which could probably soften the EoS.
In the confinement-deconfinement mechanism considered
in this paper the naturally appearing diquarks have spin-1. This
means that in the ground-state of matter they do not condensate because
of rotational invariance, unless one assumes e.g. a strong spin-spin
attraction between pairs causing (quasi) spin-less bound states.
An extension to include color superconducting  spin-zero diquarks
with use of the Hartree-Fock-Bogoliubov (HFB) theory and the impact on the
softening of the EoS is an interesting future problem.

\section{Results and discussion}

The EoSs are derived from NQMM-A and NQMM-B for 
nucleon-quark mixed matter, respectively, without and with the 
Pauli-exclusion effect for neutrons from the quark Fermi sphere.
The adjustable parameters included in these EoSs are (1) the nucleon-triquark 
coupling constant $\lambda 3$ in $V_{DQE}^{(QN)}$ and (2) the corrections 
$\Delta m_q$ for constituent quark masses:

\noindent
(1) Nucleon-triquark coupling constants are chosen as
$\lambda_3/\sqrt{4\pi}$= 0, 0.1, 0.2, 0.4, 0.8, 1.5, and
resulting $qN$ interactions are denoted as $\lambda 0$, $\lambda 01$, 
$\lambda 02$, $\lambda 04$, $\lambda 08$, $\lambda 15$, respectively.
Here $\lambda_3/\sqrt{4\pi}$=0 ($\lambda 0$) means to take only the meson-exchange 
parts $V_{EME}^{(QN)}$ without the direct-quark-exchange parts $V_{DQE}^{(QN)}$.
The limiting case of switching off all $qN$ and $qq$ interactions is denoted
as "$qt$", where only kinetic energies are taken into account in free quark states.
In Fig.\ref{MRA}, the density-dependent quark mass Eq.(\ref{mstar}) is used 
in the $\lambda15$ case, denoted as $\lambda 15'$.

\noindent
(2) Parameters  $\Delta m_q$ control onset densities for quark mixing:
In the cases of NQMM-A (NQMM-B), parameters $\Delta m_q$ are chosen so that
the onset density is $2.5 \rho_0$, $2.0 \rho_0$ and $1.6 \rho_0$ 
with $\rho_0$=0.17fm$^{-3}$, which are denoted as
NQMM A16, NQMM A20 and NQMM A25 (NQMM B16, NQMM B20 and NQMM B25), respectively.
For instance, a NQMM-A EoS for $\lambda 15$ with an onset density $2.0 \rho_0$
is denoted as NQMM A20-$\lambda 15$.

\medskip
For each $qN$ interaction, values of $\Delta m_q$ are determined
so as to give onset densities $2.5 \rho_0$, $2.0 \rho_0$ and $1.6 \rho_0$.
The determined values of $\Delta m_q$ are tabulated in Table \ref{delM}.
\begin{table}[h]
\begin{center}
\caption{Values of $\Delta m_q$ (MeV). 
}
\label{delM}
\begin{tabular}{|c|rrrrrrr|}\hline
  & $qt$ & $\lambda 0$ & $\lambda 01$ & $\lambda 02$ 
  & $\lambda 04$ & $\lambda 08$ & $\lambda 15$ \\
\hline
2.5$\rho_0$ & 15. & 10. & 25. & 40. & 65. & 120. & 200. \\
2.0$\rho_0$ & $-5.$ & $-15.$ & $-5.$ & 10. & 30. & 70. & 140. \\
1.6$\rho_0$ & $-20.$ & $-28.$ & $-20.$ & $-10.$ & 5. & 40. & 90. \\
\hline
\end{tabular}
\end{center}
\end{table}

\subsection{EoS}

\begin{figure}[h]
\begin{center}
\includegraphics*[width=8.6cm]{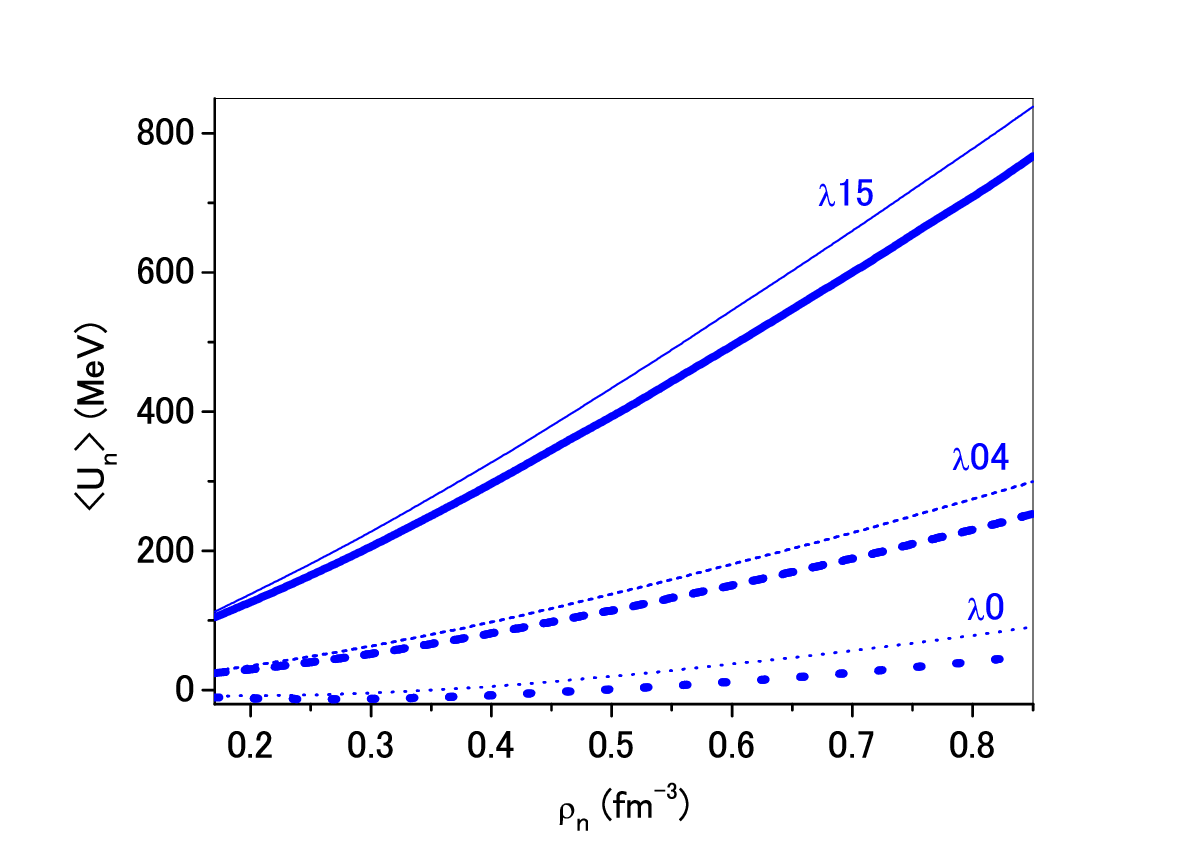}
\caption{(Color online)  Averaged potential energy of neutron $\langle U_n \rangle$ 
given by Eq.(\ref{UNQ}) as a function of neutron number density $\rho_n$. 
Dotted, short-dashed and solid curves are obtained in the cases of $\lambda 0$, 
$\lambda 04$ and $\lambda 15$, respectively,
Thick (thin) curves are in the cases of $Y_Q=$ 0.2 (0.6).}
\label{UPOT} 
\end{center}
\end{figure}

Let us demonstrate the repulsive $qN$ interactions.
In Fig.\ref{UPOT}, averaged potential energy of neutron $\langle U_n \rangle$ in 
the case of $Y_p=0$ are drawn as a function of neutron number density $\rho_n$.
$\langle U_n \rangle$ is defined by the expression
\begin{eqnarray}
\langle U_n \rangle&=& \sum_{q=u,d}
\frac{1}{\rho_q} \int_0^{k_F^q} \frac{d^3k}{(2\pi)^3} U_n^{(q)}(k)  \,
\end{eqnarray}
where $U_n^{(q)}(k)$ given by Eq.(\ref{UNQ}) are obtained from $qn$ interactions.
Dotted, short-dashed and solid curves are obtained in the cases of $qn$ interactions
$\lambda 0$, $\lambda 04$ and $\lambda 15$, respectively, 
where thick (thin) curves are in the cases of $Y_Q=$ 0.2 (0.6).
It is noted that the repulsive contributions of $\langle U_n \rangle$
by $qn$ repulsions increase rapidly with density $\rho_n$, which work 
to stiffen EoSs in high density regions.  Even in the case of $\lambda 0$,
there exists the weakly repulsive contribution from $V_{EME}^{(QN)}$.

\begin{figure}[h]
\begin{center}
\includegraphics*[width=8.6cm]{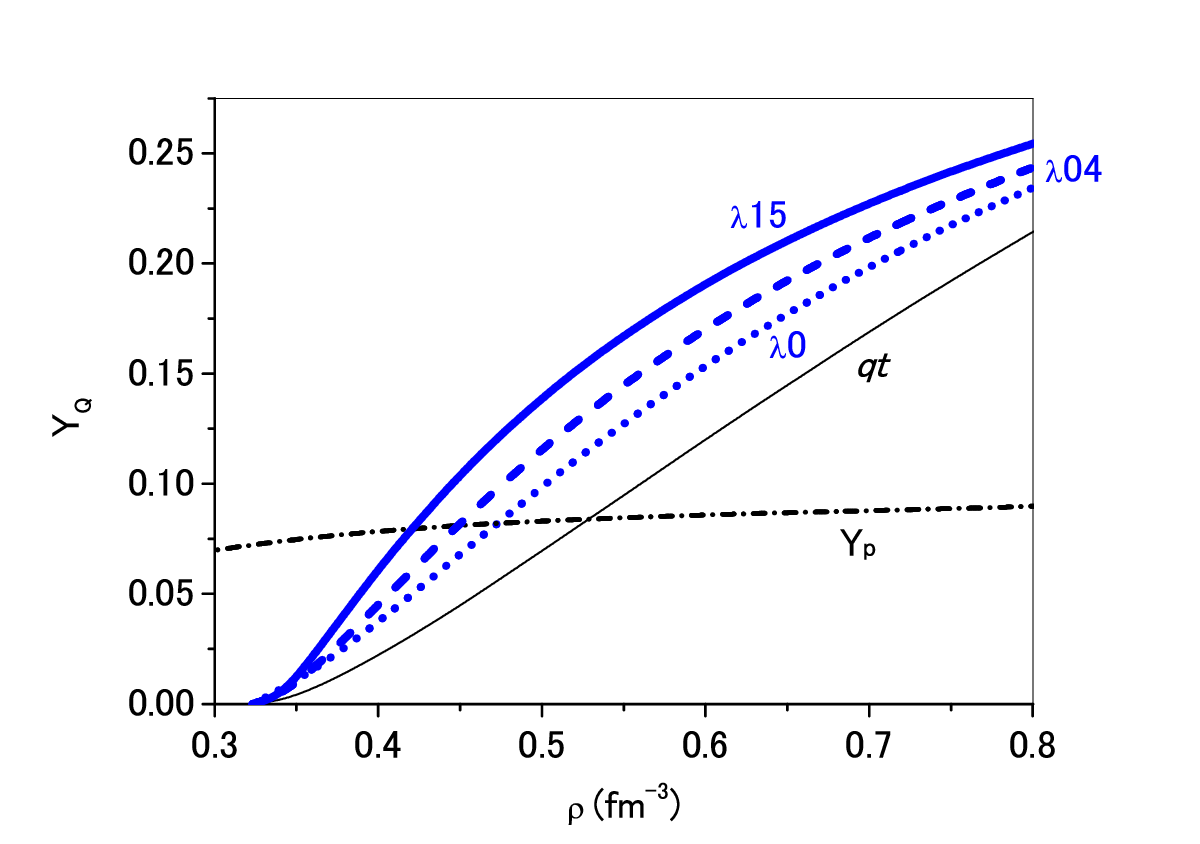}
\caption{(Color online) Quark mixing rates $Y_Q$ given as a function of total 
density $\rho=\rho_N+\rho_Q$. Solid, short-dashed, dotted and thin-solid curves
are in the cases of $\lambda 15$, $\lambda 04$, $\lambda 0$ and $qt$, respectively,
in the case of onset density $2.0 \rho_0$.
Dot-dashed curve is proton mixing rates $Y_P$ as a function of $\rho$.
}
\label{Yrho}
\end{center}
\end{figure}

\begin{figure}[h]
\begin{center}
\includegraphics*[width=8.6cm]{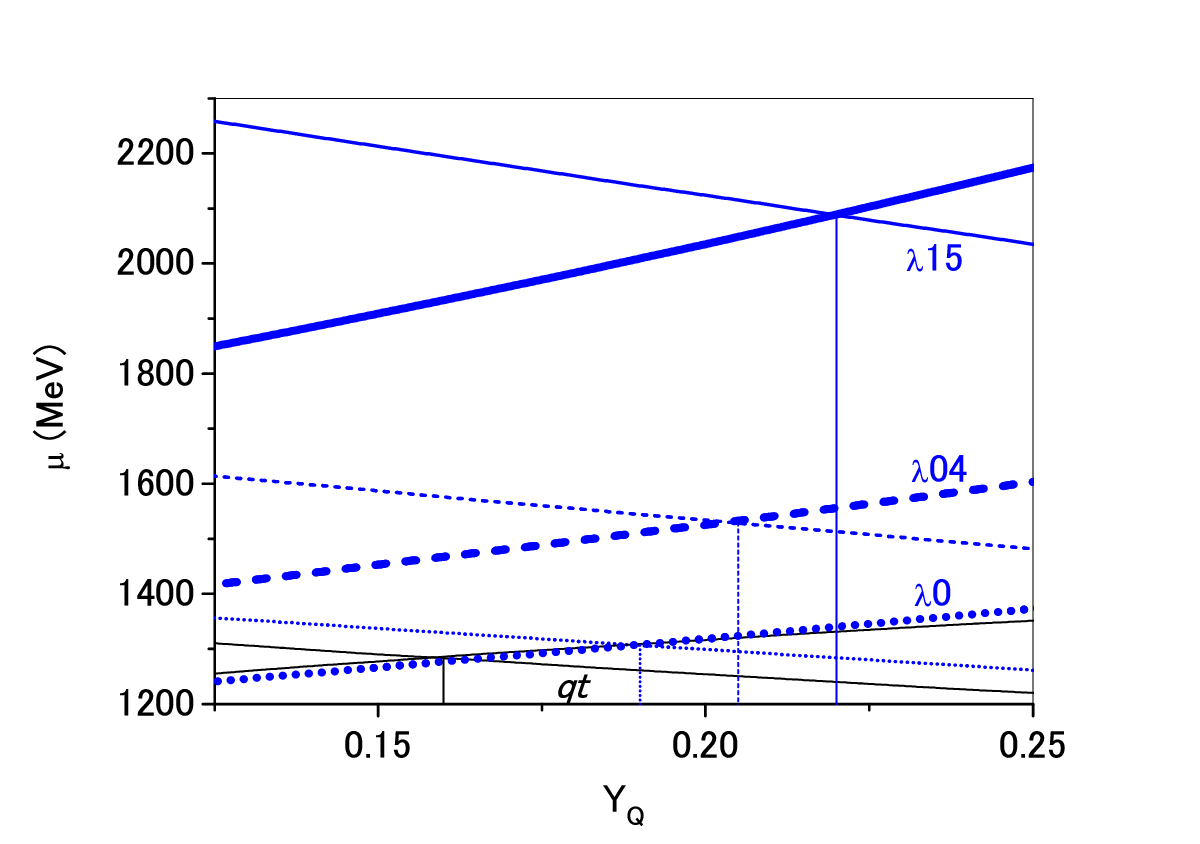}
\caption{(Color online) Neutron (quark) chemical potentials $\mu_n$
($\mu_u +2\mu_d$) as a function of $Y_Q$ in the case of 
$\rho=\rho_N+\rho_Q=4\rho_0$.  Solid, short-dashed, dotted and
thin-solid curves are in the cases of $\lambda 15$, $\lambda 04$, 
$\lambda 0$ and $qt$, respectively, in the case of onset density $2.0 \rho_0$.
Increasing and decreasing curves show $\mu_u +2\mu_d$ 
and $\mu_n$, respectively. $Y_Q$ values satisfying equilibrium conditions
($\mu_n = \mu_u +2\mu_d$) are given by abscissa values of cross points, 
where $Y_Q$ values for $\lambda 15$, $\lambda 04$, $\lambda 0$ and $qt$
are 0.22, 0.205, 0.19 and 0.16, respectively.
}
\label{MuY}
\end{center}
\end{figure}

\begin{figure}[h]
 \begin{center}
 \includegraphics*[width=8.6cm]{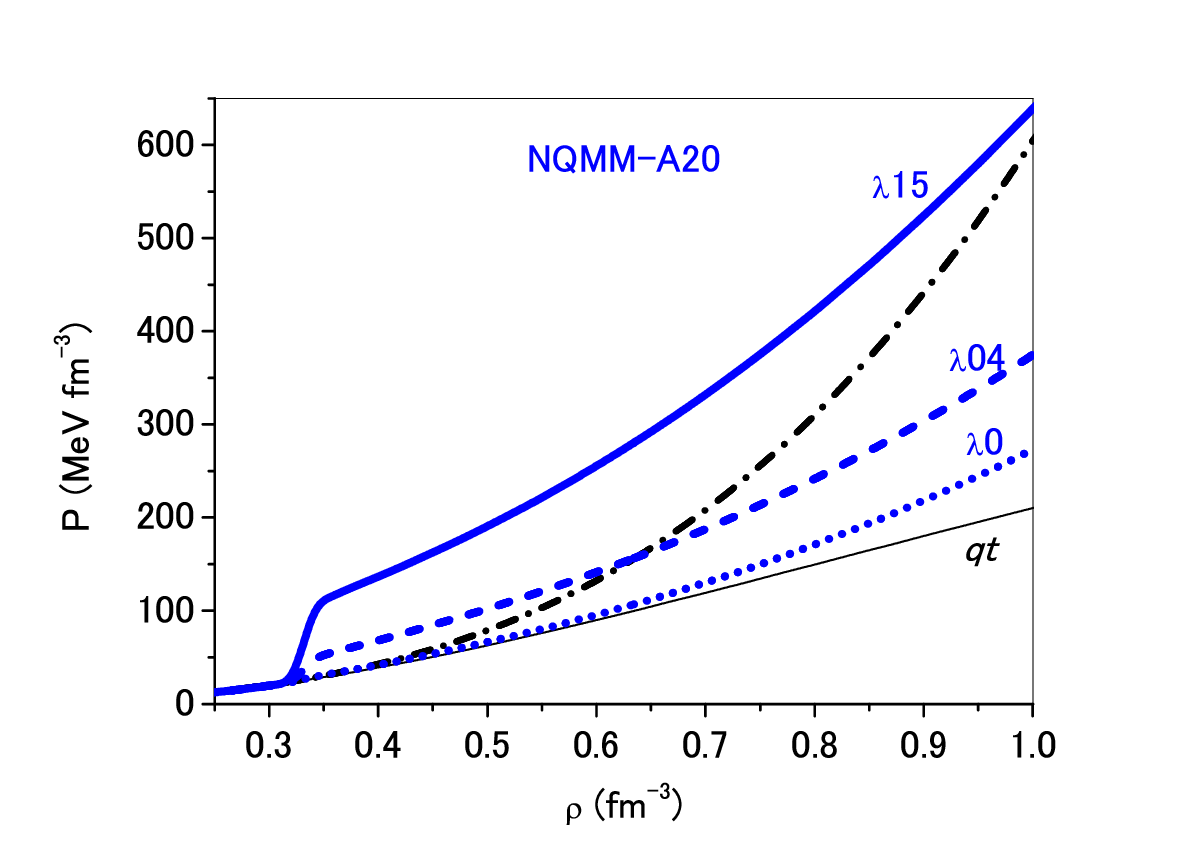}
 \caption{(Color online) Partial pressures of nucleons 
 as a function of total density $\rho=\rho_N+\rho_Q$. 
 The dot-dashed curve is pressures in nucleonic matter. 
 Solid, short-dashed, dotted and thin-solid curves are pressures for
 $\lambda 15$, $\lambda 04$, $\lambda 0$ and $qt$, respectively,
 in the case of NQMM-A20.}
 \label{PresN}
 \end{center}
 \end{figure}

 \begin{figure}[h]
 \begin{center}
 \includegraphics*[width=8.6cm]{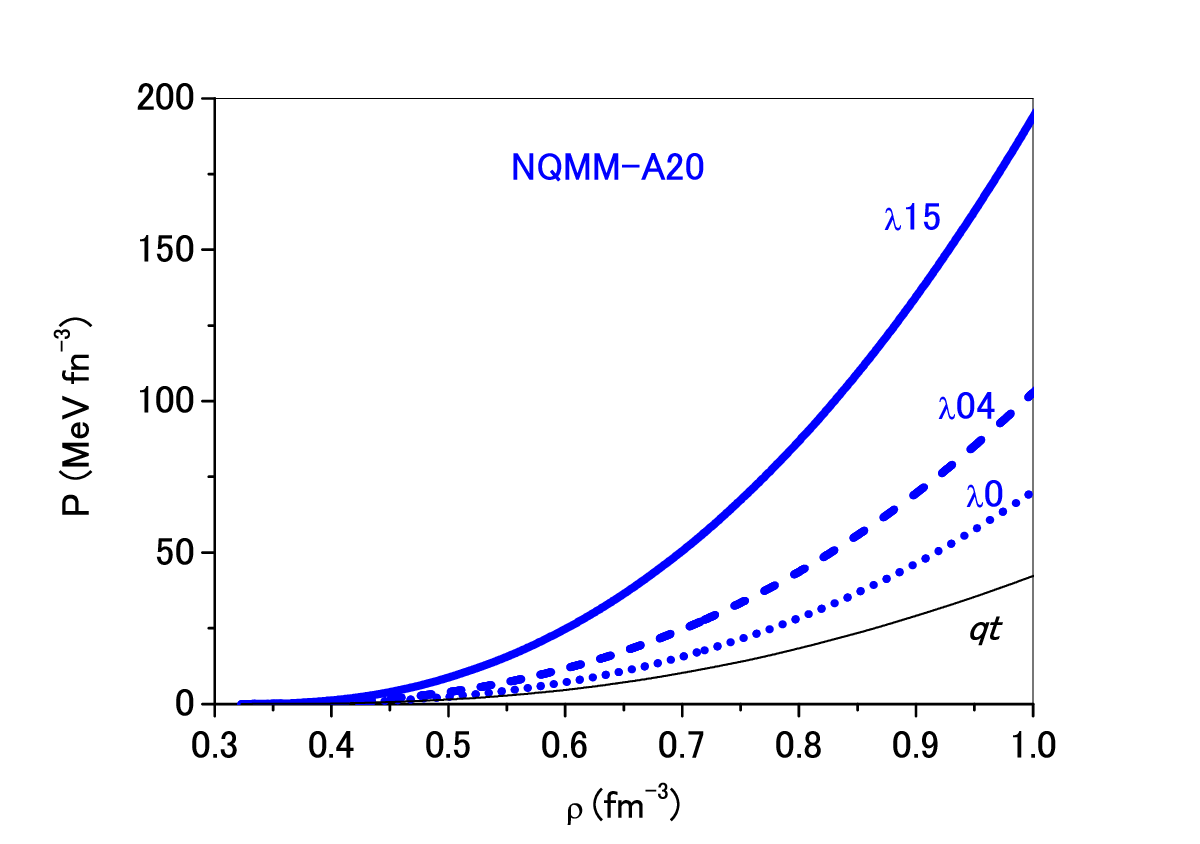}
 \caption{(Color online) Partial pressures of free quarks
 as a function of total density $\rho=\rho_N+\rho_Q$.
 Sold, short-dashed, dotted and thin-solid curves are pressures 
 for $\lambda 15$, $\lambda 04$, $\lambda 0$ and $qt$, 
 respectively, in the case of NQMM-A20.}
 \label{PresQ}
 \end{center}
 \end{figure}

In our nucleon-quark mixed matter, the quark mixing rate $Y_Q$ is
determined by the chemical equilibrium condition $\mu_n = \mu_u+2\mu_d$ 
Eq.(\ref{eq:c1}) with chemical potentials $\mu_n$, $\mu_u$ and $\mu_d$.
In Fig.\ref{Yrho}, values of $Y_Q$ are given as a function of total 
density $\rho=\rho_N+\rho_Q$. The solid, short-dashed, dotted
and thin-solid curves are in the cases of $\lambda 15$, $\lambda 04$, 
$\lambda 0$ and $qt$, respectively, in the case of onset density $2.0 \rho_0$.
In the important density region lower than about $5\rho_0$,
the $Y_Q$ values are noted to become larger, as the $qN$ interactions 
become more repulsive from $\lambda 0$ to $\lambda 15$. 
For reference, the $Y_p$ values as a function of $\rho$ are shown 
by the dot-dashed curve.

The relations between the $Y_Q$ values and the strengths of $qN$ repulsions
are demonstrated in Fig.\ref{MuY}, where the quark and neutron chemical 
potentials $\mu_u +2\mu_d$ and $\mu_n$ are drawn as a function of $Y_Q$ 
in the case of total density $\rho=\rho_N+\rho_Q=4\rho_0$.
The solid, short-dashed, dotted and thin-solid curves are in the cases of 
$\lambda 15$, $\lambda 04$, $\lambda 0$ and $qt$, respectively,
in the case of onset density $2.0 \rho_0$.
The increasing (decreasing) curves show $\mu_u +2\mu_d$ ($\mu_n$), and
the chemical equilibrium conditions $\mu_n = \mu_u +2\mu_d$ are
satisfied at cross points of increasing and decreasing curves, 
namely $Y_Q$ values satisfying equilibrium conditions are given by 
abscissa values of cross points, as indicated by vertical lines.
The $Y_Q$ values at cross points are noted to become larger
as the $qN$ interactions become more repulsive, which is the reason
why the $Y_Q$ values in Fig.\ref{Yrho} become larger as the $qN$ 
interactions become more repulsive.

In Fig.\ref{PresN} and Fig.\ref{PresQ}, partial pressures of nucleons 
and free quarks, $P_N$ and $P_Q$, are given as a function of 
total density $\rho=\rho_N+\rho_Q$, respectively.
In Fig.\ref{PresN}, the dot-dashed curve is pressures in nucleonic matter. 
The solid, short-dashed, dotted and thin-solid curves are pressures for
$\lambda 15$, $\lambda 04$, $\lambda 0$ and $qt$, respectively,
in the case of NQMM-A20. 
In Fig.\ref{PresQ}, solid, short-dashed, dotted and thin-solid curves are pressures 
$P_Q$ of free quarks for $\lambda 15$, $\lambda 04$, $\lambda 0$ and $qt$, 
respectively, in the case of NQMM-A20.
$P_Q$ values increase when the $qN$ repulsions and $Y_Q$ values increase 
from $\lambda 0$ to $\lambda 15$.  Anyway, partial pressures of free quarks 
are substantially smaller than those of neutrons.

 \begin{figure}[h]
 \begin{center}
 \includegraphics*[width=8.6cm]{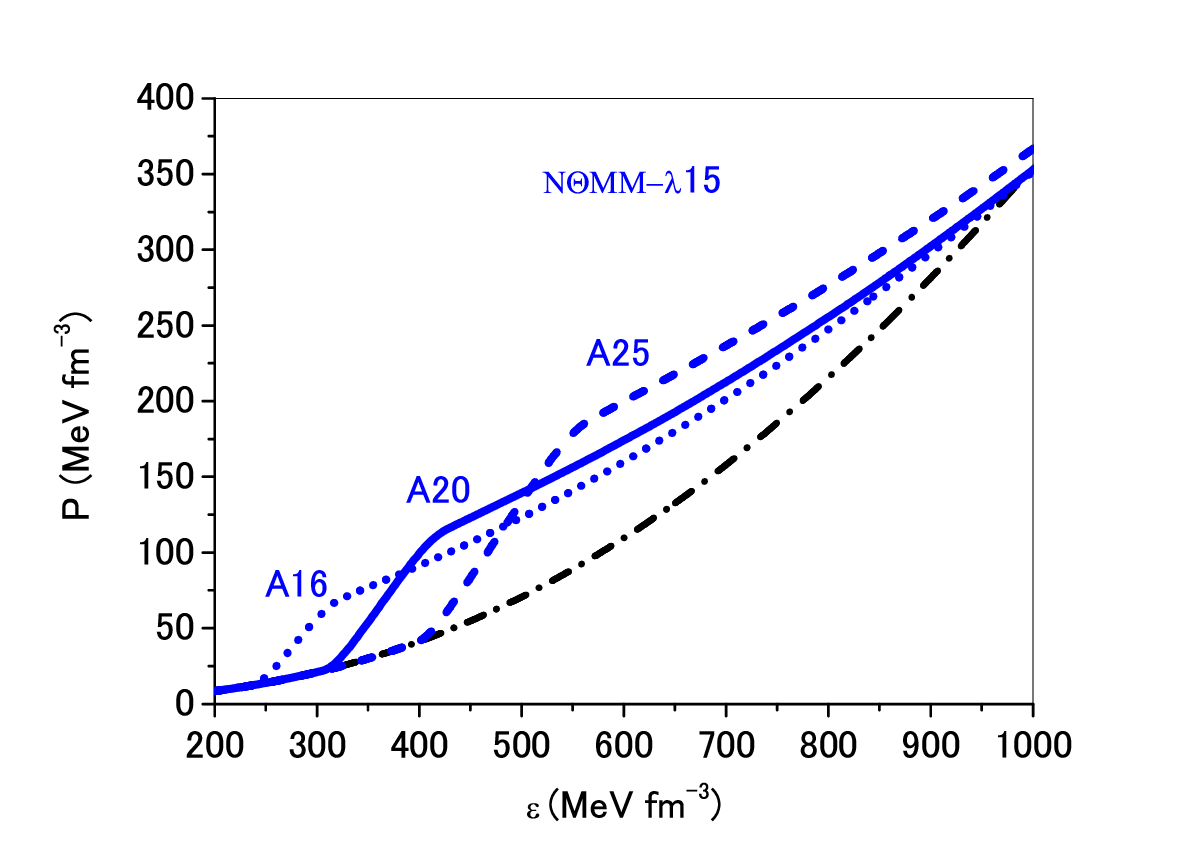}
 \caption{(Color online) Pressures as a function of energy densities
 for NQMM-$\lambda 15$.  Solid, short-dashed and dotted curves are 
 in the cases of A20, A25 and A16.
 The dot-dashed curve is pressures in nucleonic matter.
 }
 \label{EOSA1}
 \end{center}
 \end{figure}

 \begin{figure}[h]
 \begin{center}
 \includegraphics*[width=8.6cm]{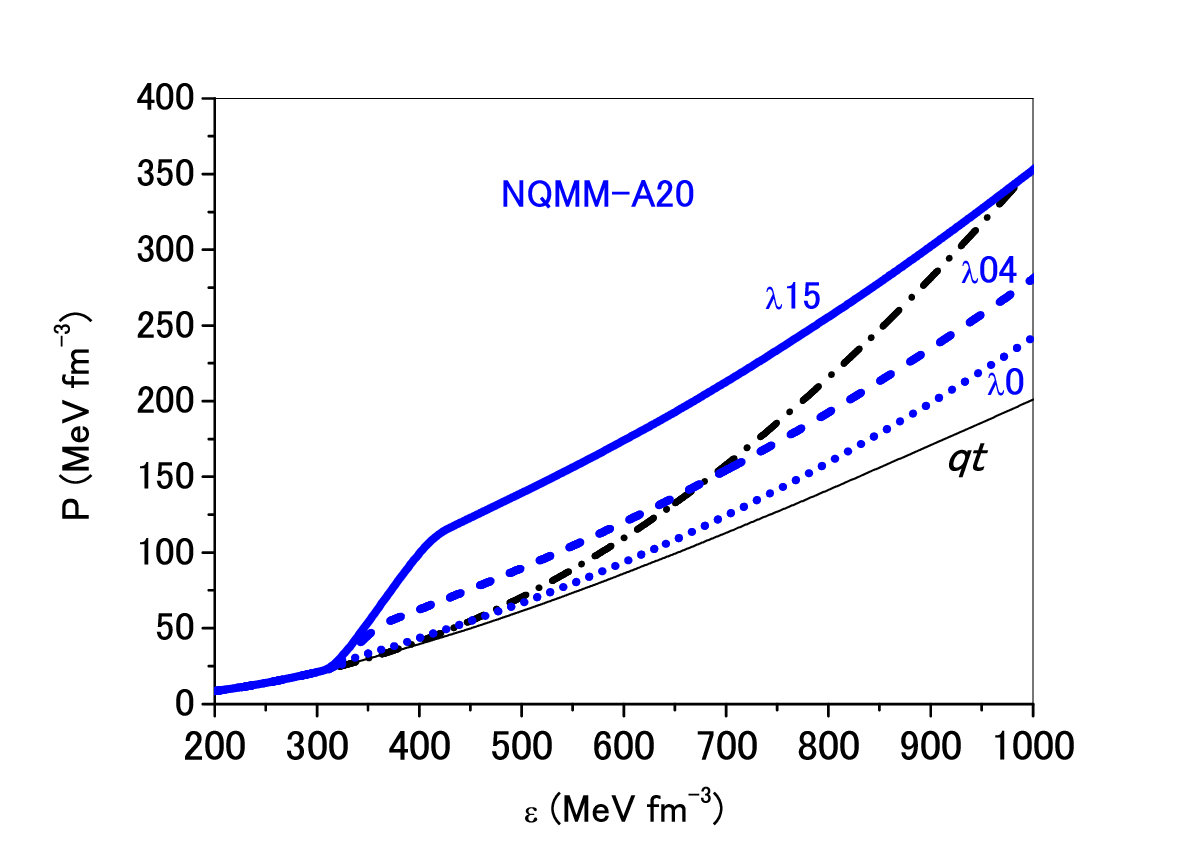}
 \caption{(Color online) Pressures as a function of energy densities
 for NQMM-A20. 
 Solid, short-dashed, dotted and thin-solid curves are in the cases 
 of $\lambda 15$, $\lambda 04$, $\lambda 0$ and $qt$, respectively.
 The dot-dashed curve is pressures in nucleonic matter. 
 }
 \label{EOSA2}
 \end{center}
 \end{figure}

In Fig.\ref{EOSA1}, pressures $P$ for NQMM-A EoSs
are drawn as a function of the energy density $\varepsilon$,
that is $P(\varepsilon)$, in the cases of $\lambda 15$ with different 
quark onset densities. The solid, short-dashed and dotted curves are 
in the cases of A20, A25 and A16, respectively.
The dot-dashed curve is pressures in nucleonic matter.
The branching points of the above three curves from this dot-dashed 
curve are related to the quark onset densities. 
Sudden increase in pressure at a quark onset point means that
the derivative $\partial P/\partial \varepsilon$ is discontinuous,
and the phase transition from nucleonic matter to nucleon-quark 
mixed matter is second-order.
As shown later, such a rapid change of pressure for energy density
produces a peak in the speed of sound \cite{KBH2022}.

In Fig.\ref{EOSA2}, pressures $P$ for NQMM-A20 EoSs are drawn as a function of
the energy density $\varepsilon$ in the case of quark onset density $2.0\rho_0$.
The dot-dashed curve is pressures in nucleonic matter. 
The solid, short-dashed, dotted and thin-solid curves are in the cases 
of $\lambda 15$, $\lambda 04$, $\lambda 0$ and $qt$, respectively.
The solid curve in this figure is the same as
the solid one in Fig.\ref{EOSA1}, both of which
are obtained by NQMM A20-$\lambda 15$.
One should note, here, the remarkable reduction of pressures from 
the dot-dashed curve for nucleonic matter to the thin-solid curve ($qt$),
which can be considered as the EoS softening caused by quark mixing
in the case of no $qN$ and $qq$ repulsions. 
In the cases of $\lambda 15$ and $\lambda 04$ with strong $qN$ di-quark 
repulsions, the EoS softening is recovered by sudden increases 
in pressure at quark onset points.
In the case of $\lambda 0$, such effects are not so noticeable.

 \begin{figure}[h]
 \begin{center}
 \includegraphics*[width=8.6cm]{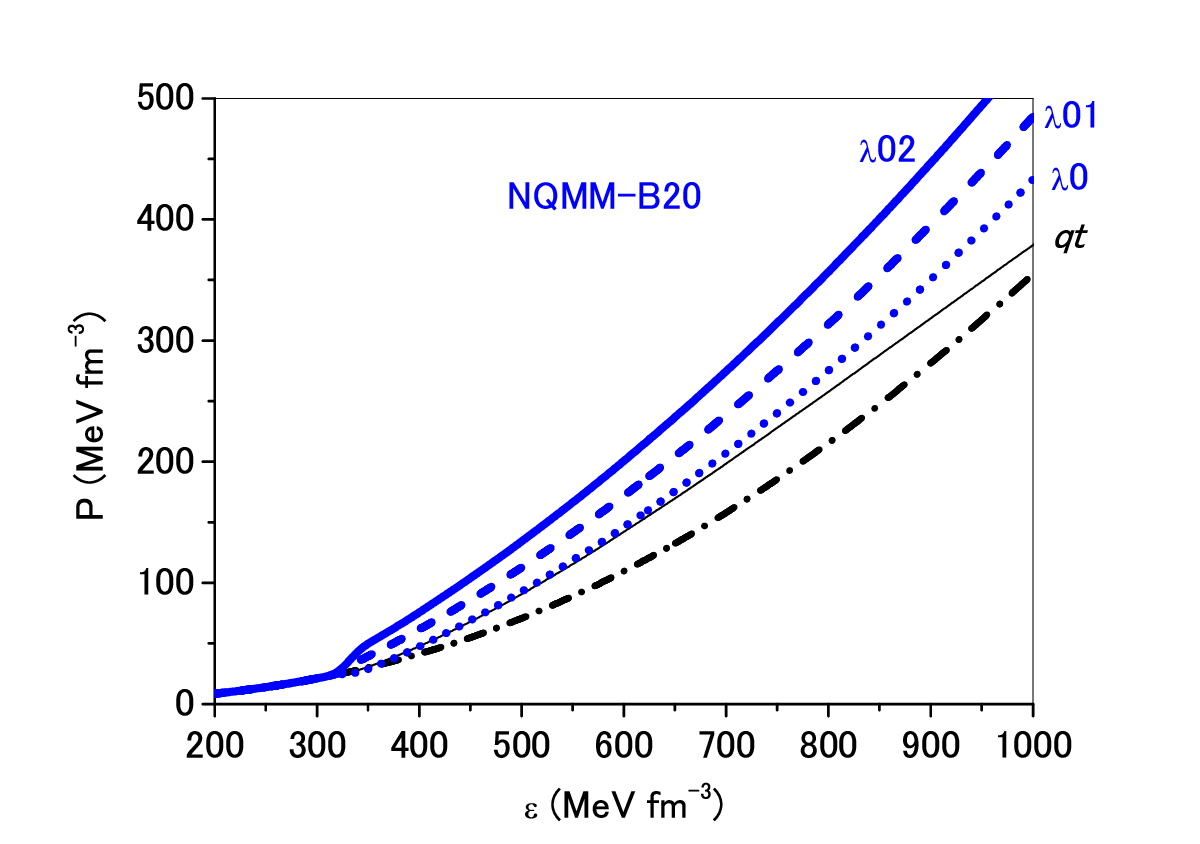}
 \caption{(Color online) Pressures as a function of energy densities
 for NQMM-B20 EoSs.
 Solid, short-dashed, dotted and thin-solid curves are in the cases 
 of $\lambda 15$, $\lambda 04$, $\lambda 0$ and $qt$, respectively.
 The dot-dashed curve is pressures in nucleonic matter.
 }
 \label{EOSB1}
 \end{center}
 \end{figure}

In Fig.\ref{EOSB1}, pressures $P$ are drawn as a function of 
the energy density $\varepsilon$ for NQMM-B20 EoSs
with quark onset density $2.0\rho_0$.
The dot-dashed curve is pressures in nucleonic matter.
The solid, short-dashed, dotted and thin-solid curves are 
obtained in the cases of $\lambda 15$, $\lambda 04$,  
$\lambda 0$ and $qt$, respectively.
These curves are noted to be above the dot-dashed curve for 
nucleonic matter. The curves for $\lambda 15$ and $\lambda 04$
are pushed up by the combined effects of the Fermi repulsions 
for nucleons from the quark Fermi sphere 
and the di-quark exchange $qN$ repulsions.

 \begin{figure}[h]
 \begin{center}
 \includegraphics*[width=8.6cm]{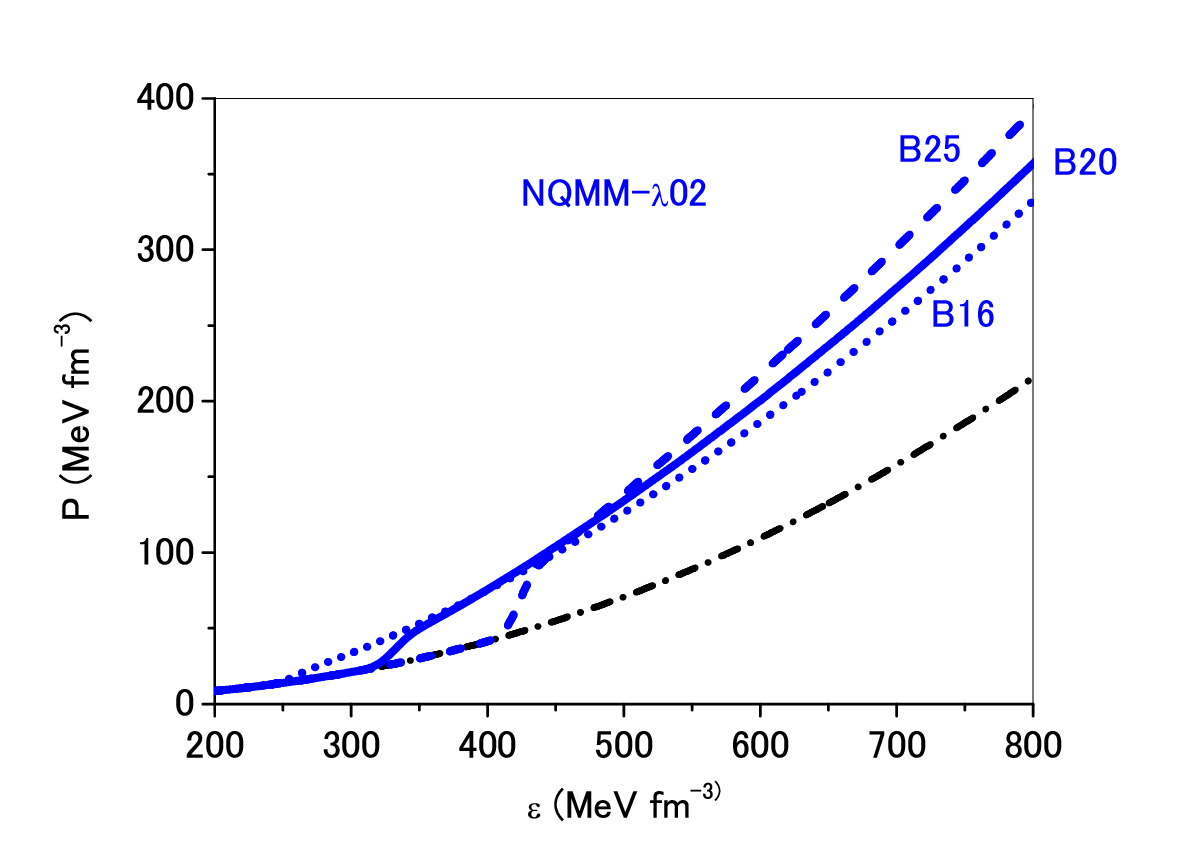}
 \caption{(Color online) Pressures as a function of energy densities
 for NQMM-$\lambda 02$.  Solid, short-dashed and dotted curves are 
 in the cases of B20, B25 and B16.
 The dot-dashed curve is pressures in nucleonic matter.
 }
 \label{EOSB2}
 \end{center}
 \end{figure}

In Fig.\ref{EOSB2}, pressures $P$ for NQMM-$\lambda 02$ EoSs
are drawn as a function of the energy density $\varepsilon$
for different quark onset densities. 
The solid, short-dashed and dotted curves are in the cases of
B20, B25 and B16, respectively. 
The solid curve in this figure is the same as
the solid one in Fig.\ref{EOSB1}, both of which
are obtained for NQMM B20-$\lambda 02$.
The dot-dashed curve is pressures in nucleonic matter.
The different quark onset densities lead to the different 
branching points from the dot-dashed curve.
Sudden increase in pressure at a quark onset point 
(second-order phase transition) is related
to rising of a $MR$ curve at a corresponding onset density,
producing a peak in the speed of sound.

\begin{figure}[h]
\begin{center}
\includegraphics*[width=8.6cm]{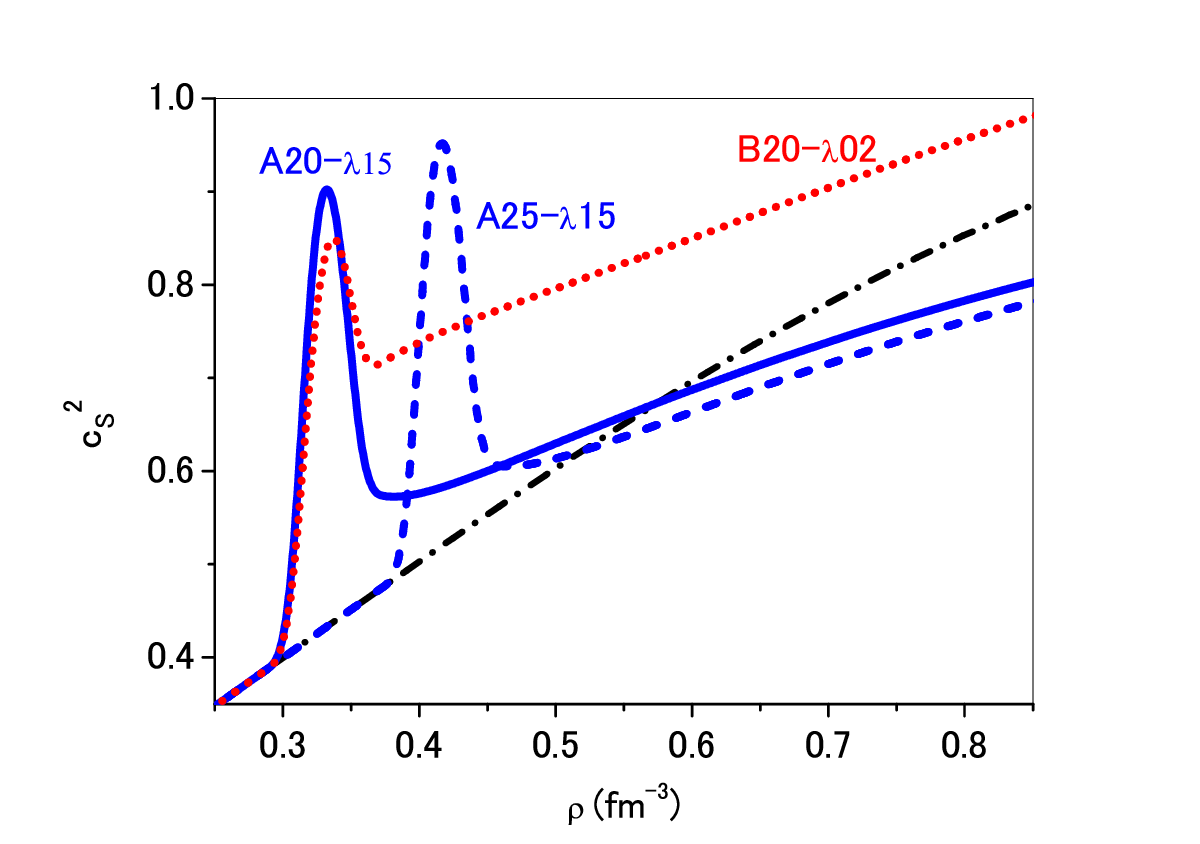}
\caption{(Color online) Sound velocities $c_s^2= \partial P/\partial \varepsilon$
are drawn as a function of baryonic density $\rho=\rho_N+\rho_Q$. 
The dot-dashed curve is sound velocities in the nucleonic matter.
Solid, short-dashed and dotted curves are sound velocities in
nucleon-quark mixed matter for NQMM A20-$\lambda 15$, A25-$\lambda 15$
and B20-$\lambda 02$, respectively.
}
\label{sound}
\end{center}
\end{figure}

\begin{figure*}[ht]
\begin{center}
 \includegraphics*[width=6.5in,height=3.4in]{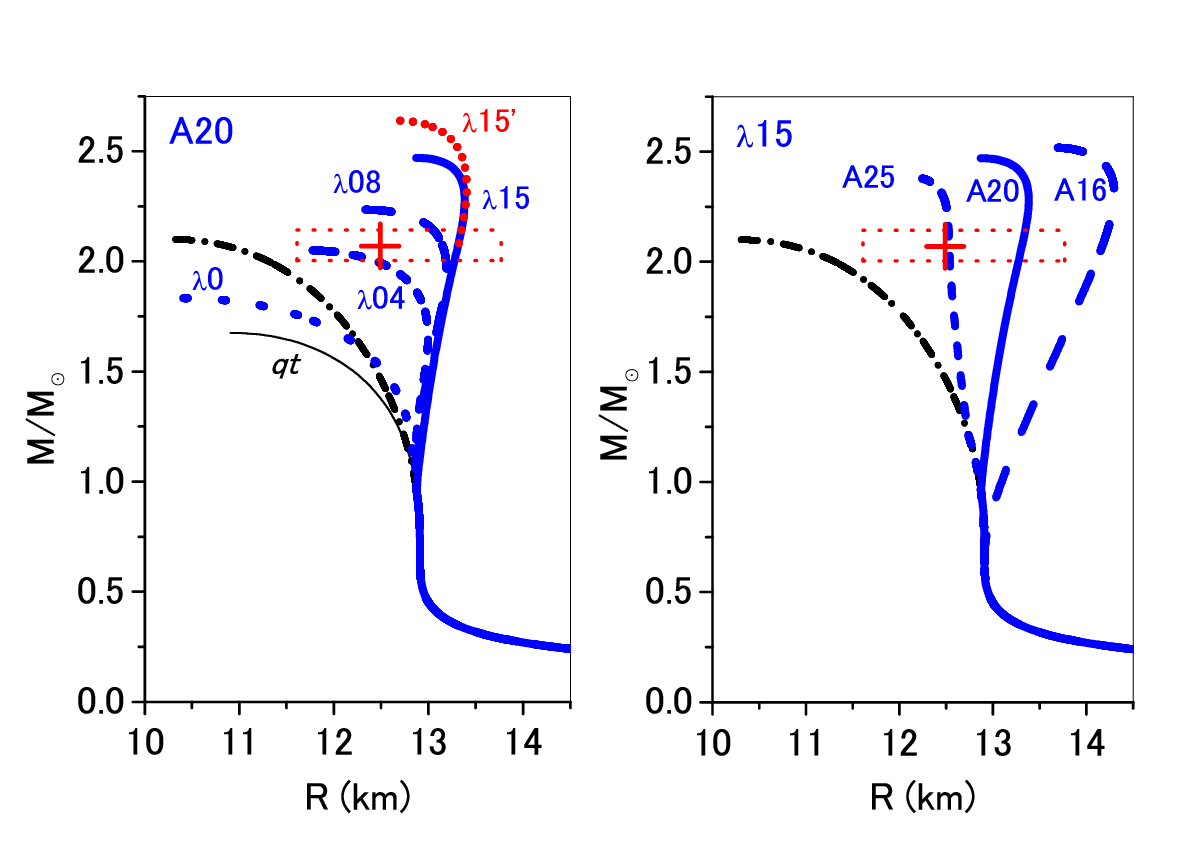}
\caption{(Color online) Star masses as a function of radius $R$ for NQMM-A EoSs. 
In the left panel, thin-solid, dotted, short-dashed, dashed, solid and short-dotted 
curves are for $qt$, $\lambda 0$, $\lambda 04$, $\lambda 08$, $\lambda 15$
and $\lambda 15'$, respectively, in the cases of the onset density $2.0\rho_0$ (A20).
In the right panel, $M(R)$ curves are for $\lambda 15$ with different quark 
onset densities. 
Solid, short-dashed and dashed curves are in the cases of onset densities 
$2.5\rho_0$ (A25), $2.0\rho_0$ (A20) and $1.6\rho_0$ (A16), respectively. 
In both panels, the dot-dashed curves are for the nucleonic matter EoS.
}
\label{MRA}
\end{center}
\end{figure*}

In Fig.\ref{sound}, sound velocities $c_s^2= \partial P/\partial \varepsilon$
in nucleon-quark mixed matter are drawn as a function of total density 
$\rho=\rho_N+\rho_Q$. The dot-dashed curve is sound velocities in the 
nucleonic matter. Solid, short-dashed and dotted curves are sound velocities 
in the cases of NQMM A20-$\lambda 15$, A25-$\lambda 15$ and B20-$\lambda 02$, 
respectively, where the peaks appear at quark onset densities $2.0\rho_0$ 
(A20 and B20) and $2.5\rho_0$ (A25).
The peak structures of sound velocities are produced by the second-order
phase transitions from nucleonic to nucleon-quark mixed states
at quark onset densities. 
This mechanism is the same as the one in \cite{MR2019}.

\begin{figure*}[ht]
\begin{center}
 \includegraphics*[width=6.5in,height=3.4in]{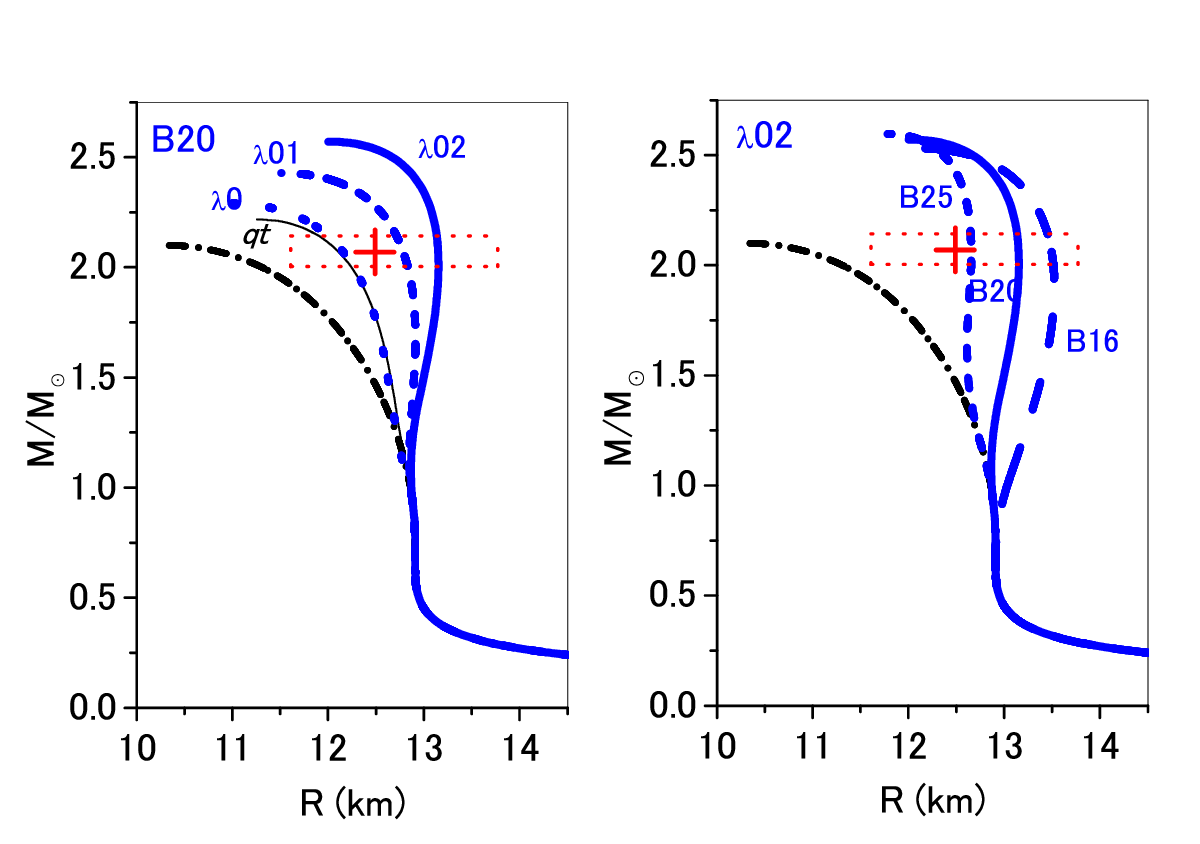}
\caption{(Color online) Star masses as a function of radius $R$ for NQMM-B EoSs
In the left panel, the thin-solid, dotted, short-dashed and solid curves 
are for $qt$, $\lambda 0$, $\lambda 01$ and $\lambda 02$, respectively.
In the right panel, $M(R)$ curves are for $\lambda 02$ with different quark 
onset densities. Solid, short-dashed and dashed curves are in the cases of 
onset densities $2.5\rho_0$ (B25), $2.0\rho_0$ (B20) and $1.6\rho_0$ (B16), 
respectively. In both panels, the dot-dashed curves are obtained by 
the nucleonic matter EoS.
}
\label{MRB}
\end{center}
\end{figure*}

\subsection{$MR$ diagrams}

We have the EoSs for several models of nucleon-quark mixed matter 
(MQMM-A and NQMM-B).
These EoSs are connected to the crust EoS \cite{Baym1,Baym2} 
at $\rho= 0.22$ fm$^{-3}$ with smooth interpolations.
Star masses $M$ as a function of radius $R$, that is $M(R)$, are 
obtained by solving the TOV equations with these EoSs 
for nucleon-quark mixed matter.

In the following figures for $M(R)$ curves, 
the regions given by $M = 2.08 \pm 0.07 M_\odot$ and $R= 12.49^{+1.28}_{-0.88}$ km
\cite{Salmi2024} are drawn by dotted rectangles, and the point  
($M=2.1M_\odot$, $R_{2.1M_\odot}$=12.5 km) is indicated by a cross symbol.
In the analysis of $M(R)$ curves, our critical guideline is that an obtained 
$M(R)$ curve reaches above this cross symbol in the $MR$ diagram.

In the right panel of Fig.\ref{MRA}, star masses are given as a function of radius 
$R$ for the NQMM-A EoSs in the cases of $\lambda 15$ with different quark onset 
densities, corresponding to the $P(\varepsilon)$ curves in Fig.\ref{EOSA1}.
The dot-dashed curves are obtained by the nucleonic matter EoS.
The solid, short-dashed and dashed curves are in the cases of
quark onset densities $2.0\rho_0$ (A20), $2.5\rho_0$ (A25) and
$1.6\rho_0$ (A16), respectively. 

In the left panel of Fig.\ref{MRA}, star masses are given as a function of 
radius $R$ for NQMM-A EoSs in the case of quark onset density $2.0\rho_0$.
The thin-solid, dotted, short-dashed, dashed and solid curves are in the 
cases of $qt$, $\lambda 0$, $\lambda 04$, $\lambda 08$ and $\lambda 15$,
respectively. The dot-dashed curve is obtained by the nucleonic matter EoS.
The thin-solid curve ($qt$) is obtained by the EoS in which all $qN$ and 
$qq$ interaction are switched of. The reason of the small maximum mass in 
this case is because of the EoS softening caused by changing of high-momentum 
neutrons at Fermi surfaces to low-momentum free quarks, which is 
the same as the mechanism of the EoS softening by hyperon mixing.
It should be noted that the $M(R)$ curves are pushed up by the 
increasing $qN$ repulsions from $\lambda 0$ to $\lambda 15$. 
The short-dotted curve above the solid curve for $\lambda 15$
is obtained by using the density-dependent quark mass Eq.(\ref{mstar}) 
in the case of $\lambda_3/\sqrt{4\pi}$=1.5 (denoted as $\lambda 15'$)
and the quark onset density $2.0\rho_0$. The $M(R)$ curve turns out 
to be pushed up further by the density-dependent effects of quark masses.

In Fig.\ref{MRB}, star masses are given as a function of radius $R$ for NQMM-B EoSs.
In the left panel, the thin-solid, dotted, short-dashed and solid curves are 
obtained for $qt$, $\lambda 0$, $\lambda 01$ and $\lambda 02$, respectively, 
with quark onset density $2.0\rho_0$.
In the right panel, they are obtained for $\lambda 02$ EoSs
with different quark onset densities.
The solid, short-dashed and dashed curves are in the cases of
quark onset densities $2.0\rho_0$ (B20), $2.5\rho_0$ (B25) and
$1.6\rho_0$ (B16), respectively. 
The dot-dashed curves in both panels are obtained by the nucleonic matter EoS.
The $M(R)$ curves (onset density $2.0\rho_0$) in the left panel 
are of good correspondence to the $P(\varepsilon)$ curves in Fig.\ref{EOSB1}.
It should be noted that all $M(R)$ curves for nucleon-quark mixed matter EoSs 
are above this dot-dashed curve differently from those in the left panel of
Fig.\ref{MRA}. Though the $M(R)$ curve for NQMM-A ($qt$) gives small maximum mass
due to the EoS softening, the curve for NQMM-B ($qt$) is pushed up 
by the Fermi repulsion for nucleons from the quark Fermi sphere 
and give maximum mass over that for the nucleonic EoS.
In Fig.\ref{MRB}, the $M(R)$ curves for $\lambda 01$ and $\lambda 02$ 
are pushed up by the combined effects of the quark Fermi repulsions 
and the increasing $qN$ di-quark exchange repulsions.

\begin{figure}[h]
\begin{center}
\includegraphics*[width=8.6cm]{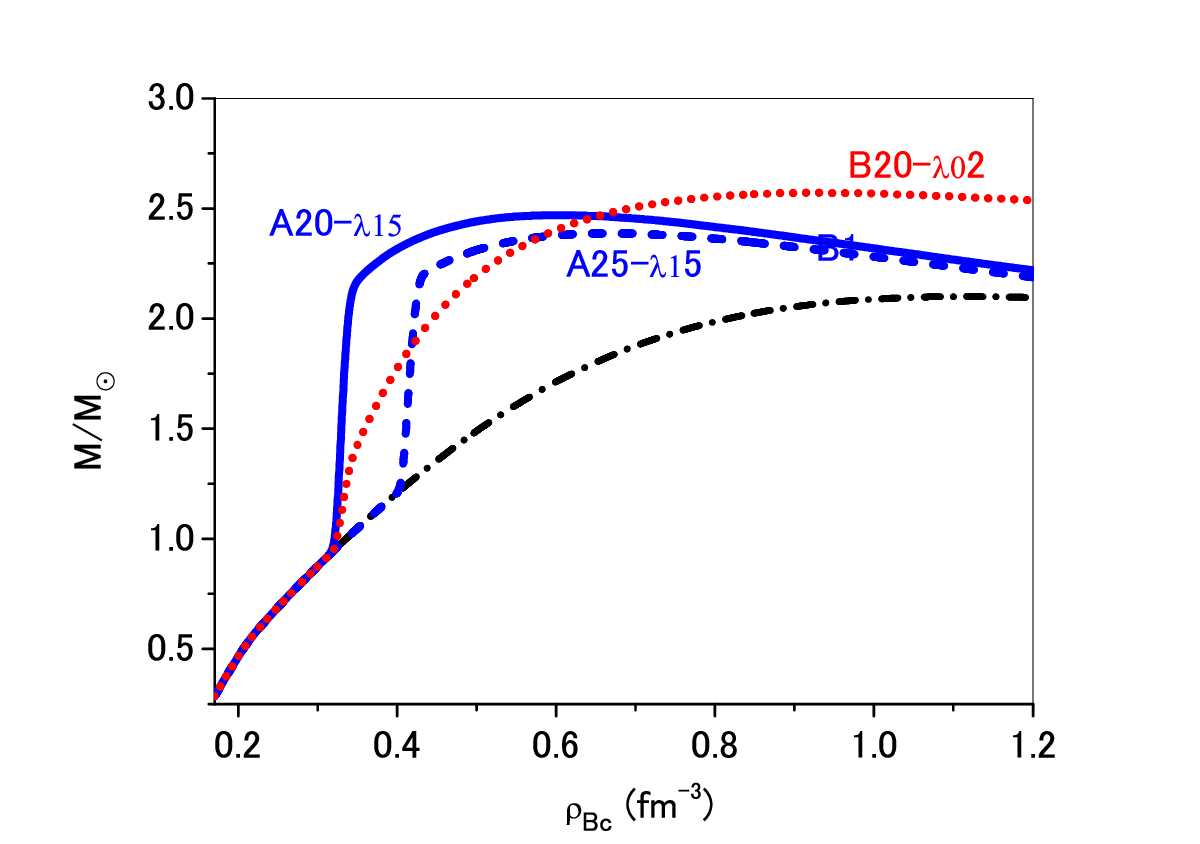}
\caption{(Color online) Star masses as a function of central baryon density $\rho_{Bc}$.
The dot-dashed curve is obtained by the nucleonic EoS. 
Solid, short-dashed and dotted curves are by NQMM A20-$\lambda 15$,
A25-$\lambda 15$ and B20-$\lambda 02$, respectively.
}
\label{Mdenc}
\end{center}
\end{figure}

In Fig.\ref{Mdenc}, star masses are given as a function 
of central baryon density $\rho_{Bc}$, that is $M(\rho_{Bc})$.
The dot-dashed curve is obtained by the nucleonic EoS.
The solid, short-dashed and dotted curves are by NQMM A20-$\lambda 15$,
A25-$\lambda 15$ and B20-$\lambda 02$, respectively.
The behaviors of $M(\rho_{Bc})$ curves respond well to the corresponding 
$P(\varepsilon)$ curves in Fig.\ref{EOSA2} and Fig.\ref{EOSB1}.
The $P(\varepsilon)$ curves, as well as the $M(\rho_{Bc})$ curves, 
are noted to be above the dot-dashed curves by the nucleonic EoS.

\begin{table}
\begin{center}
\caption{Values of maximum masses $M_{max}/M_\odot$ and
radii $R_{2.1M_\odot}$ (km) at $2.1M_\odot$ obtained for NQMM-A
and NQMM-B EoSs. NUC give the values for nucleonic matter EoS.
}
\label{MRtab}
\vskip 0.2cm
\begin{tabular}{|ll|c|c|}\hline
   &  & $M_{max}/M_\odot$ & $R_{2.1M_\odot}$ (km)  \\
\hline
A20     & $qt$          & 1.68 &        \\
        & $\lambda 0$   & 1.83 &        \\
        & $\lambda 4$   & 2.05 &        \\
        & $\lambda 8$   & 2.23 &  13.1  \\
        & $\lambda 15$  & 2.47 &  13.3  \\
        & $\lambda 15'$ & 2.64 &  13.3  \\
A25     & $\lambda 15$  & 2.37 &  12.5  \\
\hline
B20     & $qt$          & 2.22 &  12.0  \\
        & $\lambda 0$   & 2.23 &  12.1  \\
        & $\lambda 1$   & 2.43 &  12.8  \\
        & $\lambda 2$   & 2.57 &  13.1  \\
B25     & $\lambda 2$   & 2.60 &  12.7  \\
\hline
NUC     &     & 2.10 &  11.3  \\
\hline
\end{tabular}
\end{center}
\end{table}

Table \ref{MRtab} summarize the values of maximum masses $M_{max}/M_\odot$ 
and radii $R_{2.1M_\odot}$ (km) found in Fig.\ref{MRA} and Fig.\ref{MRB}, 
where the values of $R_{2.1M_\odot}$ turn out to be consistent with 
the observed data $R= 12.49^{+1.28}_{-0.88}$ km \cite{Salmi2024}.

\bigskip

\section{Conclusion}
The nucleon-quark mixed matter is defined in the BHF framework,
which is suitable to treat $qN$ and $qq$ interactions without being
obscured by ad hoc parameters such as thickness $\Delta$ of neutron 
Fermi layer used in the quarkyonic matter \cite{MR2019}. 
In this framework the quark density in nucleon-quark mixed matter is 
determined by the equilibrium conditions between chemical potentials 
of nucleons and free quarks. Our $qN$ interaction is
composed of meson-exchange and di-quark exchange potentials,
playing critical roles for our nucleon-quark mixed matter EoS.
The di-quark potential is strongly repulsive and the nucleon-triquark coupling
constant plays a role as a critical parameter for the stiffness of our EoS.
In the nucleon-quark mixed matter, Fermi spheres of nucleons 
and free quarks are occupied by nucleons with momenta of $0<k<k_F^N$
and quarks with momenta of $0<k<k_F^q$, respectively,
from which NQMM-A EoSs are derived straightforwardly.
Taking the basic concept of quakyonic matter \cite{MR2019} into account,
the nucleon Fermi levels are pushed up to the one with momenta of 
$k_0^N<k<k_1^N$ (Fermi layer) by the Fermi exclusion effect for nucleons
from the quark Fermi sphere. The NQMM-B EoSs are derived by incorporating 
this quark Fermi repulsion into our nucleon-quark mixed matter. 

Our EoSs are controlled basically by the strength of di-quark exchange
repulsion and the effective quark mass: The former and latter determine
the stiffness of EoS and the quark onset density, respectively.
Then, the phase transition from nucleonic matter to nucleon-quark 
mixed matter is second-order, and there appear peak structures 
in sound velocities at quark onset densities.

The features of our nucleon-quark mixed matter are demonstrated
by pressures as a function of energy density $P(\varepsilon)$
and neutron-star mass-radius curves $M(R)$. 
In the cases of NQMM-A EoSs, the $P(\varepsilon)$ and $M(R)$
curves for nucleon-quark mixed matter are pushed up above those
for nucleonic matter by the $qN$ repulsive interactions, 
as found in Fig.\ref{EOSA2} and Fig.\ref{MRA}.
These pushing-up effects are further enhanced by taking account
of density dependences of quark masses.
In the limiting case of switching off all $qN$ and $qq$ interactions, 
the $M(R)$ curves are far below the curves for nucleonic matter, 
which can be considered as remarkable EoS softening caused by changing 
of high-momentum neutrons to low-momentum free quarks.
As $qN$ repulsions increase, the $MR$ curves for nucleon-quark mixed 
matter are pushed above the point ($M=2.1M_\odot$, $R_{2.1M_\odot}$=12.5 km) 
indicated by the recent radius observation of the massive neutron stars.
In our nucleon-quark mixed matter different from the quakyonic matter 
\cite{MR2019}, these pushing-up effects for $MR$ curves are realized 
by the strong $qN$ repulsions in NQMM-A EoSs, without taking account 
of the Fermi exclusion effect for nucleons from the quark 
Fermi sphere included in NQMM-B EoSs.

In the cases of NQMM-B EoSs, the $P(\varepsilon)$ and $M(R)$ curves 
for nucleon-quark mixed matter are always above those for nucleonic 
matter, as found in Fig.\ref{EOSB1} and Fig.\ref{MRB}, respectively.
Even if the $qN$ repulsions are switched off, the $MR$ curves 
come near the point ($M=2.1M_\odot$, $R_{2.1M_\odot}$=12.5 km)  
owing to the Fermi repulsion for nucleons from the quark Fermi sphere.
In the case of NQMM-B EoSs, the pushing-up effects for $MR$ curves
are realized by additive contributions from the $qN$ repulsions
and the Fermi exclusion effect from the quark Fermi sphere.

Our approach to neutron-star EoSs is based on the nucleon-quark 
mixed matter in the BHF framework, in which the $qN$ di-quark exchange repulsions
play important roles. The derived EoSs are consistent with observed data 
of $M_{max}$ and $R_{2M_\odot}$.
In relation to the ``hyperon puzzle", there still remains the issue of 
the size of an EoS softening induced by hyperon/$s$-quark mixing 
in the nucleon/baryon-quark mixed matter.
The extension to include hyperons and strange-quarks is a natural step
in the further development of this work.
Furthermore, an extension by incorporating color superconductivity 
spin-zero pairs using the HFB theory, studying e.g. the impact 
on the softening of the EoS is interesting.

 \onecolumngrid

\section*{Acknowledgments}
The author (N.Y.) was supported by JSPS KAKENHI Grant Number 24K07054.

 \appendix 

 \vspace*{2mm} 
 \section{ Di-quark exchange Interaction}                                                  
 \label{app:DQEXCH}
\noindent The di-quark exchange potential $V_{DQE}^{(qN)}$ is derived from a description of the 
confinement-deconfinement process at high baryon-densities via the 
(density dependent) nucleon-triquark coupling 
\begin{eqnarray}
	{\cal L}^{(1)}_{int} = -\lambda_3 \bigl[\bar{\psi}(x) \eta_N(x) 
	+ \bar{\eta}_N(x) \psi(x)\bigr] 
\label{eq:Ldeconf1a} \end{eqnarray}
	with \cite{Ioffe81}
\begin{eqnarray}
	\eta_N(x) = \left[\widetilde{q}^a(x)C\gamma^\mu q^b(x)\right] \gamma_5\gamma_\mu q^c(x) f^{abc},
\end{eqnarray}
where C is the charge conjugation operator, and momentarily we left out the isospin labels.\\
The interaction Lagrangian in (\ref{eq:Ldeconf1a}) with the tri-quark field $\eta_N(x)$ 
is rewritten using di-quark fields for two reasons:  
(i) the functional form of the partition function is difficult to handle,    
and (ii) di-quarks are meaningful physical entities.
In terms of the (bosonic) di-quark fields $\chi_\mu^a(x)$ 
\begin{subequations}
\begin{eqnarray}
\eta_N(x) &=& (\hbar c)^2 \gamma_5\gamma^\mu q^a(x)\cdot \chi_\mu^a(x), \\
	\chi_\mu^a(x) &\equiv& f^{abc} \widetilde{q}^b(x)C\gamma_\mu q^c(x)/(\hbar c)^2.
\end{eqnarray}
\end{subequations}
The interaction (\ref{eq:Ldeconf1a}) becomes
\begin{eqnarray}
	{\cal L}^{(1)}_{int}  = -\lambda_3 (\hbar c)^2 \bigl[
\left(\bar{\psi}(x)\gamma_5\gamma^\mu q^a(x)\right) \chi^a_\mu(x) + h.c. \bigr]  
\label{eq:Ldeconf2a} \end{eqnarray}

Using the grand-canonical partition functional ${\cal Z}_G$ description of matter 
and treating $\chi_\mu^a(x)$ within 
the auxiliary field method \cite{Bender77} by introducing the $\chi_\mu^a(x)$-field 
via the Lagrangian
\begin{equation}
	{\cal L}_\chi = \bar{\lambda}_3^2\biggl\{\chi^{a \dagger}_\mu(x) \chi^{\mu a}(x) 
	- \bigl[\chi_\mu^{a \dagger}(x) 
	\bigl(\tilde{q}^b(x)C\gamma^\mu q^c(x)\bigr) f^{abc} + {\it h.c.}\bigr]\biggr\}
\end{equation}

the partition functional becomes, see \cite{Kap89},
\begin{eqnarray}
	Z_G &=& \int [d\bar{\psi}] [d\psi] [d\bar{q}] [dq] [d\sigma] [d\omega_\mu]
	\int {\cal D}\chi^{a \mu} {\cal D}{\chi}_\mu^{a \dagger}
	\exp \biggl[ \int_0^\beta d\tau \int d^3x \cdot\nonumber\\ && \times
	 \bigl( {\cal L}_N + {\cal L}_Q 
	+ {\cal L}_M + {\cal L}_{\chi}   
	+\mu_N \psi^\dagger \psi + \mu_Q q^\dagger q \bigr)\biggr].
\label{eq:Mix.63}\end{eqnarray}

For example in the Walecka-model \cite{Walecka74}    
\begin{subequations}\label{eq:Mix.34}
\begin{eqnarray}
{\cal L}_Q &\rightarrow& \bar{q}(x)\biggl[ i\gamma_\mu\left(\partial^\mu+\frac{i}{3}g_\omega \omega^\mu\right)
-\bigl(m_Q-\frac{1}{3}g_\sigma \sigma \bigr)\biggr]\ q(x) 
\\  
{\cal L}_N &\rightarrow& \bar{\psi}(x)\biggl[ i\gamma_\mu\left(\partial^\mu+ig_\omega \omega^\mu\right)
-\bigl(m_N-g_\sigma \sigma  \bigr)\biggr]\ \psi(x) 
\\  
{\cal L}_M &\rightarrow& 
+\frac{1}{2}\left(\partial^\mu\sigma \partial_\mu \sigma-m_\sigma^2 \sigma^2\right)
-\frac{1}{4}\omega_{\mu\nu}\omega^{\mu\nu}+\frac{1}{2}m_\omega^2 \omega_\mu \omega^\mu    
\end{eqnarray} \end{subequations}
The terms in ${\cal L}_\chi$ are schematically, apart from an overall
	factor $\bar{\lambda}_3^2$,
\begin{eqnarray}
	{\cal L}_\chi \sim 
&& \chi_\mu^\dagger\chi^\mu-\chi_\mu^\dagger B^\mu-B_\mu^\dagger \chi^\mu =
 (\chi_\mu-B_\mu)^\dagger (\chi^\mu-B^\mu)-B_\mu^\dagger B^\mu, 
\label{eq:Mix.36d}\end{eqnarray}
with $B_\mu = \left[\bar{\psi}(x)\gamma_5\gamma_\mu q(x)\right]/(\hbar c)^2$.   
The integration over the shifted $\left(\chi_\mu^a-B_\mu^a\right)$ 
di-quark fields gives in the exponential of $Z_G$ the term
\begin{eqnarray}
	&& -\bar{\lambda}_3^2\ B_\mu^\dagger B^\mu = 
-\bar{\lambda}_3^2(\bar{\psi}\gamma_5\gamma_\mu q)(\bar{q}\gamma_5\gamma^\mu \psi) 
\label{eq:Mix.36e}\end{eqnarray}
	where $\bar{\lambda}_3 \equiv \lambda_3/(\hbar c)$.
The Lagrangian in ${\cal Z}_G$  becomes 
	${\cal L}= {\cal L}_N + {\cal L}_Q+ {\cal L}_M +{\cal L}^{(2)}_{int}$
with 
\begin{eqnarray}
	{\cal L}^{(2)}_{int} &=& - \bar{\lambda}_3^2\ B_\mu^\dagger B^\mu = 
	-\bar{\lambda}_3^2\ \bigl(\bar{\psi}\gamma_5\gamma_\mu q\bigr) 
\bigl(\bar{q}\gamma_5\gamma^\mu\psi\bigr),    
\label{eq:Mix.179}\end{eqnarray}
which is the interaction in (\ref{eq:Ldeconf3}).

 \twocolumngrid


\begin{thebibliography}{99}

\bibitem{Demorest10}
P.B. Demorest, T. Pennucci, S.M. Ransom, M.S.E. Roberts, and J.W. Hessels,
Nature (London) {\bf 467}, 1081 (2010). 

\bibitem{Antoniadis13}
J. Antoniadis {\it et al.},
Science {\bf 340}, 6131 (2013).

\bibitem{Cromartie2020}
H.T. Cromartie {\it et al.},
Nat. Astron. {\bf 4}, 72 (2020).

\bibitem{Romani2022}
R.W. Romani {\it et al.},
Astrophysical J. Lett. 934:L17 (2022).

\bibitem{Drago}
A. Drago, A. Lavagno, G. Pagliara, and D. Pigato,
Phys. Rev. C{\bf 90}, 065809 (2014).

\bibitem{Kaplan}
D.B. Kaplan, A.E. Nelson,
Phys. Lett. B{\bf 175}, 57 (1986); {\bf 179}, 409 (1986).

\bibitem{Brown}
G.E. Brown, C.-H. Lee, M. Rho, and V. Thorsson,
Nucl. Phys. A{\bf 567}, 937 (1994).

\bibitem{Thorsson}
V. Thorsson, M. Prakash, and J.M. Lattimer,
Nucl. Phys. A{\bf 572}, 693 (1994).

\bibitem{Lee}
C.-H. Lee,
Phys. Rep. {\bf 275}, 255 (1996).

\bibitem{GleSch}
N.K. Glendenning and J. Schaffner-Bielich,
Phys. Rev. Lett. {\bf 81}, 4564 (1998).

\bibitem{Baldo2006}
M.Baldo, G.F. Burgio, and H-J. Schulze,
{\it Superdense QCD Matter and Compact stars 
(NATO Science Series 2: Mathematics, Physics and Chemistry vol.197)}
(2006) ed D.Blaschke and D. Sedrakian.

\bibitem{Ozel}
F. \"{O}zel, D. Psaltis, S. Ransom, P. Demorest, and M. Alford,
Astrophys. J. Lett. {\bf 724}, L199 (2010).

\bibitem{Weissenborn}
I. Sagert, G. Pagliara, M, Hempel, and Schaffner-Bielich,
Astrophys. J. Lett. {\bf 740}, L14 (2011).

\bibitem{Klahn}
T. K\"{a}hn, D. Blaschke, and D. Lastowiecki,
Phys. Rev. D{\bf 88}, 085001 (2013).

\bibitem{Bonanno}
L. Bonanno and A. Sedrakian,
Astrophys. J{\bf 539}, 416 (2012).

\bibitem{Lastowiecki2012}
R. Lastowiecki, D. Blaschke, H. Grigorian, and S. Typel,
Acta Phys. Pol. {\bf B5}, 535 (2012).

\bibitem{Shahrbaf1}
M. Shahrbaf, D. Blaschke, A.G. Grunfeld, and H.R. Moshfegh,
Phys. Rev. {\bf C101}, 025807 (2020).

\bibitem{Shahrbaf2}
M. Shahrbaf, D. Blaschke, and S. Khanmohamadi,
J. Phys. G: Nucl. Part. Phys. {\bf 47}, 115201 (2020).

\bibitem{Otto2020}
K. Otto, M. Oertel, and B-J Schaefer,
Phys. Rev. {\bf D101}, 103021 (2020); 
arXiv:1910.11929 (2020)

\bibitem{KBH2022}
T. Kojo, G. Baym, and T. Hatsuda,
Astrophys. J. {\bf 934}, 46 (2022),
arXiv:2111.11919 (2022).

\bibitem{YYR2022}
Y. Yamamoto, N. Yasutake, and Th. A. Rijken,
Phys. Rev. {\bf C105}, 015804 (2022). 

\bibitem{YYR2023}
Y. Yamamoto, N. Yasutake, and Th. A. Rijken,
Phys. Rev. {\bf C108}, 035811 (2023). 

\bibitem{Miller2021}
M.C. Miller, et al..
Astrophys. J. Lett. {\bf 918}, L28 (2021),
arXiv:2105.06979 [astro-ph.HE]

\bibitem{Riley2021}
T.E. Riley, et al.. 
Astrophys. J. Lett. {\bf 918}, L27 (2021),
arXiv:2105.06980 [astro-ph.HE]

\bibitem{Legred2021}
I. Legred, K. Chatziioannou, R. Essick, S. Han, and P. Landy,
Phys. Rev. D {\bf 104}. 063003 (2021),

\bibitem{Salmi2024}
T. Salmi, et al.. 
arXiv:2406.14466 [astro-ph.HE]

\bibitem{Dittmann2024}
A.J. Dittmann, et al.. 
arXiv:2406.14467 [astro-ph.HE]

\bibitem{NYT}
S. Nishizaki, Y. Yamamoto, and T. Takatsuka,
Prog. Theor. Phys. {\bf105}, 607 (2001); {\bf 108}, 703 (2002).

\bibitem{Vidana11}
I. Vida\^{n}a,D. Logoteta, C. Provid\^{e}ncia, A. Polls, and I. Bombaci,
Eur.Phys.Lett. {\bf 94}, 1002 (2011).

\bibitem{YFYR14}
Y. Yamamoto, T. Furumoto. Yasutake, and Th.A Rijken,
Phys Rev. C {bf 90} 045805 (2014).

\bibitem{YFYR16}
Y. Yamamoto, T. Furumoto, N. Yasutake, and Th.A Rijken,
Eur. Phys. J. A{\bf 52}, 19 (2016).

\bibitem{YTTFYR17}
Y. Yamamoto, H. Togashi, T. Tamagawa. T. Furumoto, N. Yasutake, and Th.A. Rijken,
Phys. Rev. C{\bf 96}, 065804 (2017).

\bibitem{Lonardoni}
D. Lonardoni A. Lovato, S. Gandolfi, and F. Pederiva,
Phys. Rev. Lett. {\bf 114}, 092301 (2015).

\bibitem{Logoteta}
D. Logoteta, I. Vida\^{n}a, I. Bombaci,
Eur. Phys. J. A{\bf 55}, 207 (2019).

\bibitem{Cerstung}
D. Cerstung, N. Kaiser. and W. Weise,
Eur. Phys. J. A{\bf 56}, 175 (2020).

\bibitem{ESC16I}
M. M. Nagels, Th. A. Rijken, and Y. Yamamoto,
Phys. Rev. {\bf C99}, 044002 (2019).

\bibitem{ESC16II}
M. M. Nagels, Th. A. Rijken, and Y. Yamamoto,
Phys. Rev. {\bf C99}, 044003 (2019).

\bibitem{ESC16III}
M. M. Nagels, Th. A. Rijken, and Y. Yamamoto,
Phys. Rev. {\bf C102}, 054003 (2020).

\bibitem{MP2007}
L. McLerran and H.D. Pisarski,
Nucl. Phys. A{\bf 796}, 83 (2007).

\bibitem{HMP2008}
Y. Hidaka, L. McLerran, and H.D. Pisarski,
Nucl. Phys. A{\bf 808}, 117 (2008).

\bibitem{MR2019}
L. McLerran and S. Reddy,
Phys. Rev. Lett. {\bf 122}, 122701 (2019).

\bibitem{HMLCP2019}
S. Han, M.A.A. Mamun, S. Lalit, C. Constantinou, and M. Prakash,
Phys. Rev. D{\bf 100}, 103022 (2019).

\bibitem{DHJ2020}
D.C. Duarte, S. Hernande-Ortiz, and K.S. Jeong,
Phps. Rev. C{\bf 102}, 025203 (2020); 065202 (2020).

\bibitem{ZL2020}
T. Zhao and J.M. Lattimer,
Phys. Rev. {\bf D102}, 023021 (2020). 

\bibitem{MHPC2021}
J. Margueron, H. Hansen, and P. Proust, and G.Chanfray,
Phys. Rev. {\bf C104}, 055803 (2021). 

\bibitem{Cao2022}
G. Cao
Phys. Rev. {\bf D105}, 114020 (2022).

\bibitem{APR98}
A. Akmal, V.R. Pandharipande, and D.G. Ravenhall,
Phys. Rev. C{\bf 58}, 1804 (1998). 


\bibitem{Togashi1}
H. Togashi and M. Takano,
Nucl. Phys. A{\bf 902}, 53 (2013).

\bibitem{Rijken24b}
Th.A.\ Rijken and Y.\ Yamamoto, {\it Quark-Quark and Quark-Nucleon potential Model.
ESC Meson-exchange Interactions},  THEF-NIJM 24.02, \\
{\it\bf http://nn-online.org/eprints} 2024. 

\bibitem{Rijken24a}
Th.A.\ Rijken, {\it Quark-Nucleon Di-quark-exchange Potentials},
THEF-NIJM 24.03, \\
{\it\bf http://nn-online.org/eprints} 2024.

\bibitem{Ioffe81}
B.L.\ Ioffe, Nucl.\ Phys,\ {\bf B 188} (1981) 317;
B.L.\ Ioffe and A.V.\ Smilga, Nucl.\ Phys.\ {\bf B 232} (1984) 109.

\bibitem{Bender77}
C.M. Bender and F.\ Cooper, Ann.\ of Phys. (N.Y.) {\bf 109}, 165 (1977).

\bibitem{Kap89} 
J.I.\ Kapusta and C.\ Gale, "Finite-Temperature Field Theory. Principles and
Applications", Cambridge University Press, Cambridge, England (1989),  
Chapter 11.
\bibitem{Walecka74}
J.D.\ Walecka, Annals of Physics (N.Y.) {\bf 83}, 491 (1974).







\bibitem{YasMar}
N. Yasutake and T. Maruyama,
Phys. Rev. D{\bf 109}, 043056 (2024). 

\bibitem{Baym1}
G. Baym, A. Bethe, and C. Pethick,
Nucl. Phys. A{\bf 175}, 225 (1971).

\bibitem{Baym2}
G. Baym, C.J. Pethick, and P. Sutherland,
Astrophys. J.{\bf 170}, 299 (1971).


\end{thebibliography}
\end{document}